\pdfoutput=1
\documentclass[sigconf]{acmart}
\usepackage{multirow}
\usepackage{makecell}
\usepackage{bbding}
\usepackage{diagbox}
\usepackage{mathrsfs}
\usepackage{color}
\usepackage{colortbl}
\usepackage{enumitem}
\usepackage{algorithm}
\usepackage{algpseudocode}
\usepackage{tcolorbox}
\usepackage{graphicx} 
\usepackage{wrapfig}
\usepackage{caption}
\definecolor{skyblue}{RGB}{135,206,235} 
\definecolor{lightblue}{RGB}{235,245,255}
\usepackage{subcaption}
\usepackage{balance}
\usepackage{colortbl}
\usepackage{xcolor} 
\usepackage{makecell} 
\usepackage{amsmath}

\usepackage{amssymb}
\usepackage{booktabs}
\usepackage{enumitem}
\usepackage{multirow}
\AtBeginDocument{%
  }

\setcopyright{cc}
\setcctype{by}
\copyrightyear{2027}
\acmYear{2027}
\acmDOI{10.1145/3862322.3863404}
\acmConference[WSDM '27]{The Twentieth ACM International Conference on Web Search and Data Mining}{February 15--19, 2027}{Hong Kong, Hong Kong}
\acmBooktitle{Proceedings of the Twentieth ACM International Conference on Web Search and Data Mining (WSDM '27), February 15--19, 2027, Hong Kong, Hong Kong}
\acmISBN{978-1-4503-XXXX-X/2027/02}

\begin{document}

\title{RACE: Ranking-Uncertainty Aware Capacity-Adaptive Customization for Efficient Sequential Recommendation} \author{Tianqi Liu}\orcid{0009-0002-7217-1290}\affiliation{\institution{Zhejiang University}\city{Hangzhou}\state{Zhejiang}\country{China}}\email{tqliu@zju.edu.cn} \author{Kaipu Sun}\orcid{0009-0007-1432-5905}\affiliation{\institution{Zhejiang University}\city{Hangzhou}\state{Zhejiang}\country{China}}\email{sunkp@zju.edu.cn} \author{Kairui Fu}\orcid{0009-0004-0284-671X}\affiliation{\institution{Zhejiang University}\city{Hangzhou}\state{Zhejiang}\country{China}}\email{fukairui.fkr@zju.edu.cn} \author{Wenkai Wang}\orcid{0009-0007-1148-4169}\affiliation{\institution{Zhejiang University}\city{Hangzhou}\state{Zhejiang}\country{China}}\email{22551237@zju.edu.cn} \author{Sihao Liu}\orcid{0009-0001-8369-7290}\affiliation{\institution{Zhejiang University}\city{Hangzhou}\state{Zhejiang}\country{China}}\email{lyosihao@zju.edu.cn} \author{Hui Zang}\orcid{0009-0003-5775-8648}\affiliation{\institution{Huawei Technologies Co., Ltd.}\city{Hangzhou}\state{Zhejiang}\country{China}}\email{zanghui@huawei.com} \author{Ran Chen}\orcid{0000-0002-2770-4070}\affiliation{\institution{Huawei Technologies Co., Ltd.}\city{Hangzhou}\state{Zhejiang}\country{China}}\email{chenran50@huawei.com} \author{Tao Wang}\orcid{0009-0006-6319-3075}\affiliation{\institution{Huawei Technologies Co., Ltd.}\city{Hangzhou}\state{Zhejiang}\country{China}}\email{wangtao155@huawei.com} \author{Shengyu Zhang}\authornote{Corresponding author.}\orcid{0000-0002-0030-8289}\affiliation{\institution{Zhejiang University}\city{Hangzhou}\state{Zhejiang}\country{China}}\email{sy\_zhang@zju.edu.cn} \renewcommand{\shortauthors}{Liu et al.}

\begin{abstract} Efficient sequential recommendation has emerged as a promising paradigm for providing real-time personalized service.
To deploy models under resource-constrained environment, model compression becomes a necessity, yet
existing methods typically apply uniform compression ratios, overlooking the heterogeneity in user intent complexity.
This "one-size-fits-all" approach leads to a fundamental \textbf{Capacity-Complexity Mismatch}, where computational resources are either wasted on simple instances or insufficient for complex ones. To address this, we propose \textbf{RACE} (\textbf{R}anking-uncertainty \textbf{A}ware \textbf{C}apacity-Adaptive Customization for \textbf{E}fficient Sequential Recommendation), a framework that dynamically adapts both model structures and input sequences with uncertainty-aware compression ratios. To ensure capacity-adaptive inference, 
RACE captures the model's intrinsic uncertainty of its predictions and transforms it  
into a proactive resource allocator. 
Specifically, we derive a list-wise ranking-uncertainty metric from the distributional stability of top-$K$ predictions, which serves as a measurable proxy for instance complexity.
A lightweight predictor is then trained via self-supervision to anticipate this uncertainty from early-layer features. Guided by this signal, RACE performs instance-specific slimming in two dimensions: a hypernetwork-style channel sensitivity extractor identifies parameter redundancy, while importance-aware progressive sequence filtering mitigates data redundancy. Extensive experiments on four popular datasets demonstrate that RACE outperforms state-of-the-art baselines on accuracy, efficiency, and adaptability.

\end{abstract}



\maketitle
\section{Introduction}

In recent years, sequential recommendation has witnessed remarkable progress driven by the advancement of deep learning~\cite{lei2021semi,zhao2019leveraging,dai2024modeling,cai2021category,thinkrec2026}. Modern approaches~\cite{kang2018self,10.1145/3357384.3357895,shin2024attentive,zhou2020s3,forge2026,rankgr2026,slide2026} leverage increasingly sophisticated Transformer-based architectures to capture user preference patterns from interaction histories.
The prevailing methods deploy models on cloud servers, where centralized 
computational resources support serving billions of users.
Although effective, this approach requires uploading user interaction data to remote servers~\cite{zhou2018deep}, which raises both network latency issues and privacy concerns~\cite{yin2024device,gong2020edgerec}. Moreover, maintaining personalized models for individual users is computationally prohibitive at scale~\cite{khan2019edge}.

\begin{figure*}[t]
    \centering \includegraphics[width=0.98\linewidth]{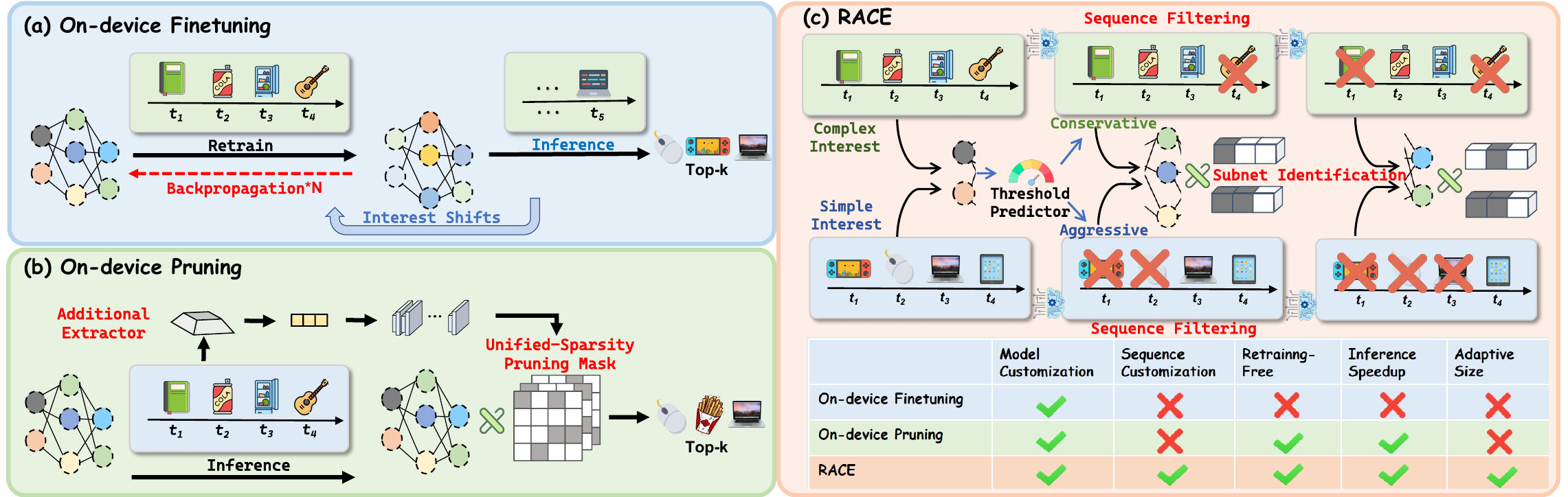}
    \vspace{-0.5mm}
    \caption{Comparison of on-device recommendation paradigms. 
(a) \textbf{On-device Finetuning} incurs computational overhead due to backpropagation, and often falls short for capturing shifting interests. 
(b) \textbf{On-device Pruning} applies a unified-sparsity mask and relies on additional extractors, unable to adapt to the complexity of user intents. 
(c) Our proposed \textbf{RACE} framework introduces a \textit{Ranking-Uncertainty Driven} mechanism to dynamically adapt model size to instance complexity. In comparison, RACE achieves sequence and structure slimming and customization in an intrinsic feature reuse and retraining-free manner.}
\label{fig:intro}
\end{figure*}

To address these challenges, efficient sequential recommendation has gained increasing attention, leveraging advances in mobile computing to support real-time reranking~\cite{wang2020convergence,gong2020edgerec,gong2022real,xi2023device}. While this paradigm mitigates latency and 
privacy concerns~\cite{khan2019edge, yin2024device}, achieving personalization on resource-constrained devices remains challenging. Some prior works adapt models to individual users through on-device 
finetuning~\cite{10.1007/s00778-021-00700-6,yan2022device,mairittha2020improving}. However, as depicted in Figure~\ref{fig:intro}(a), the required backpropagation imposes significant computational burdens. Additionally, in sustainable deployment scenarios~\cite{yin2024device}, the time-consuming update process forces systems to operate with stale models whenever user interests shift~\cite{du2023idnp,wang2020time,chang2017streaming}, degrading recommendation quality. Other works pursue resource-aware model customization~\cite{khan2019edge,liu2025chord}. 
These methods mainly rely on compression techniques: network pruning removes redundancy by identifying sparse masks~\cite{lee2018snip,sreenivasan2022rare,fu2024DIET} (Figure~\ref{fig:intro}(b)), while weight decomposition~\cite{xia2022device,yan2022apg} approximates dense matrices via low-rank factorization. Despite their effectiveness, these compression-based techniques apply \textit{static} model and input size constraints, overlooking the heterogeneity of user intent complexity~\cite{yin2024device,wang2020convergence}. 

In reality, the complexity of user preferences varies significantly across \textit{instances} (i.e., the most recent interaction sequences)~\cite{shao2025dtrec}. 
For users with stable and explicit interests, a relatively lightweight model may suffice to capture the pattern, while those with volatile or complex interaction histories demand greater model expressiveness.
This suggests that a static "one-size-fits-all" architecture can lead to a \textbf{Capacity-Complexity Mismatch}, where resources are either wasted on simple instances or insufficient for hard ones. Achieving capacity-adaptive inference thus becomes essential, yet two critical challenges remain: 

\begin{tcolorbox}[
  colback=gray!4, colframe=black!40, boxrule=0.4pt, arc=2pt,
  left=6pt, right=6pt, top=4pt, bottom=4pt,
  before skip=6pt, after skip=6pt
]
\itshape
\textbf{Challenge (i):} How to decide the ideal capacity allocation strategy?
\textbf{Challenge (ii):} How to achieve personalized slimming under the allocated capacity?
\end{tcolorbox}

\begin{figure}[h]
    \centering
    \vspace{2mm}
        \includegraphics[width=0.92\linewidth]{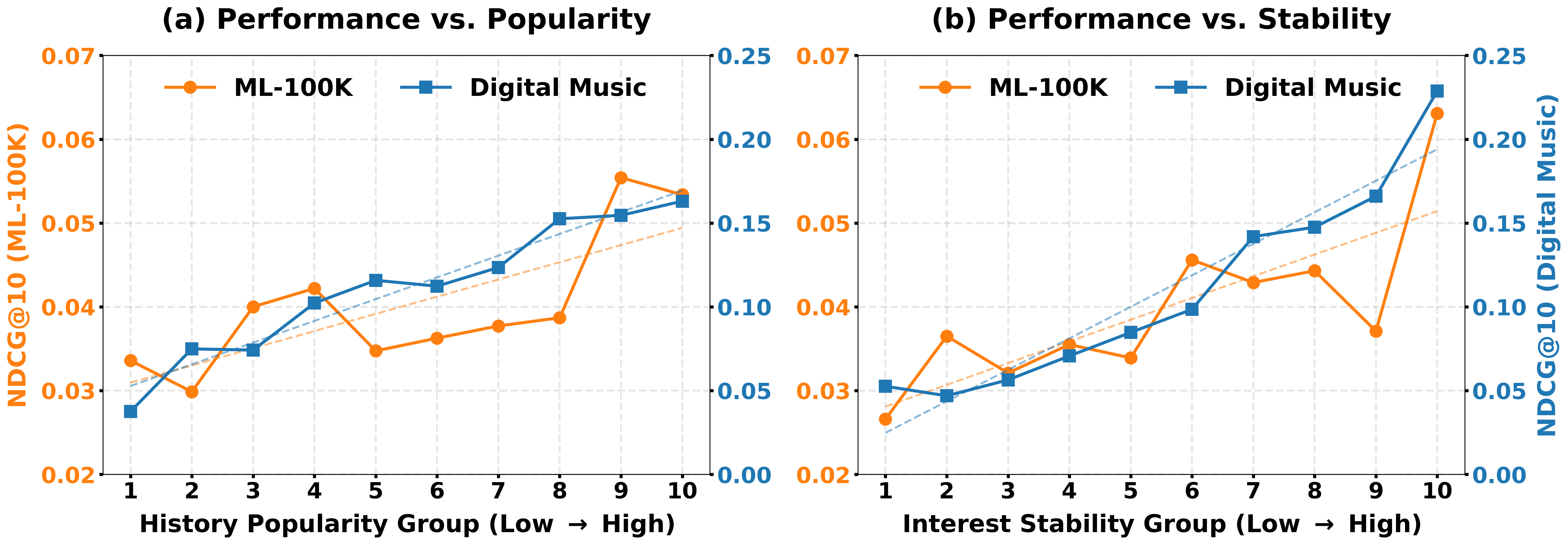}
        \vspace{-1mm}
    \caption{Correlation analysis between model performance and two input-side heuristics: (a) history popularity; (b) interest stability. Neither reliably indicates instance complexity. }
    \label{fig:race_pop_sta}
    \vspace{-1mm}
\end{figure}
The first challenge demands a \textit{resource allocation metric} that reflects instance complexity. A naive solution might rely on intuitive heuristics such as user history popularity or user interest stability. Unfortunately, as Figure~\ref{fig:race_pop_sta} shows, neither metric correlates clearly with model performance, suggesting that instance complexity is not a static input feature, but a dynamic property reflecting the confidence of the model itself. Thus, accurately capturing it requires measuring the \textbf{model's internal uncertainty}.
However, existing model uncertainty methods~\cite{heusspredictive2023,yangmitigating2023} operate primarily at the point-wise level, quantifying prediction variance for individual items while ignoring interdependencies among candidates. 
Since ranking tasks fundamentally concern relative ordering rather than absolute scores, capturing uncertainty from a \textbf{list-wise perspective} becomes essential. Moreover, to guide capacity allocation before full inference, such uncertainty should be anticipated from early-layer features rather than computed post-hoc.

The second challenge lies in determining \textit{what to slim} under the allocated 
capacity. Due to interest heterogeneity~\cite{qian2022intelligent,lv2024intelligent,yao2021device,tan2021sparse}, parameter and data sensitivity vary across instances, motivating us to jointly optimize both \textbf{parameter space} (model structure) and \textbf{data space} (input sequences) in an interest-aware manner. To identify redundancy, existing methods often incur additional feature-extraction modules~\cite{fu2024DIET}, or extra neural-networks for prediction~\cite{ye2021tr,fu2024DIET,guan2022transkimmer}. Although effective, they come with an additional computational burden. We instead seek to reuse intrinsic features and utilize lightweight analysis modules to achieve accuracy improvement and inference speedup with  minimal additional cost.

In light of this, we propose \textbf{RACE}, a \textbf{R}anking-uncertainty \textbf{A}ware \textbf{C}apacity-Adaptive Customization framework for \textbf{E}fficient sequential recommendation. As illustrated in Figure~\ref{fig:intro}(c), RACE establishes a capacity-adaptive customization and slimming for both sequences and model structures without on-device backpropagation. 
To estimate the model's inner confidence, we design a \textbf{list-wise ranking-uncertainty metric} that quantifies the structural volatility of predicted lists by jointly modeling the distributional characteristics (means and variances) of top-$k$ predictions via a negative log-likelihood formulation. We further demonstrate that this metric exhibits \textit{a clear negative correlation} with recommendation performance, validating it as a principled proxy for instance complexity. Crucially, RACE transforms this uncertainty from a post-hoc 
performance indicator into a \textbf{proactive resource allocator} through a self-supervised learning mechanism. A lightweight predictor then learns to anticipate 
ranking-uncertainty from early-layer features, enabling the model to autonomously calibrate its computational budget to instance complexity (\textbf{Challenge i}).

Driven by this uncertainty-aware threshold, RACE executes multi-granularity customization across both parameter and data spaces. Specifically, in the parameter space, we identify \textbf{interest-compatible sub-networks} adhering to an intrinsic feature reuse manner. RACE repurposes the backbone's hidden states as a dynamic user embedding, 
which a hypernetwork-style sensitivity extractor then consumes to assess
channel sensitivity, enabling personalized channel deactivation at negligible cost. Simultaneously, in the data space, the uncertainty-aware signal governs \textbf{progressive sequence filtering}, which calculates attention-aware relative importance and adaptively retains semantically core interactions while suppressing noisy ones at each block (\textbf{Challenge ii}).

We conduct extensive experiments on four real-world datasets and two representative backbones, SASRec~\cite{kang2018self} and BSARec~\cite{shin2024attentive}. Our main contributions are summarized as follows:

\begin{itemize}[leftmargin=*,nosep]
\item We reveal the capacity-complexity mismatch in on-device recommendation and propose RACE, a framework that utilizes uncertainty to adapt computational resources to instance complexity.

\item We establish a ranking-uncertainty driven resource allocation paradigm
that transforms list-wise ranking stability into a proactive signal for adaptive 
inference via self-supervised anticipation.

\item We design multi-granularity customization mechanisms, including hypernetwork-style sub-network identification and progressive sequence filtering, achieving joint optimization of model and data redundancy in a single forward pass.

\item Extensive experiments demonstrate that RACE consistently outperforms state-of-the-art baselines in both accuracy and efficiency while exhibiting strong adaptability.
\end{itemize}

\section{Related works}

\subsection{Efficient Sequential Recommendation}
Deploying models under resource-constrained conditions necessitates model compression~\cite{gong2020edgerec,yin2024device,khan2019edge} on the model side or the input side. On the model side, following the lottery ticket hypothesis, SNIP~\cite{lee2018snip} and GEM~\cite{sreenivasan2022rare} identify sparse subnetworks before or at the beginning of training, while PEMN~\cite{bai2022parameter} learns binary masks over frozen random weights. STTD~\cite{xia2022device} and APG~\cite{yan2022apg} instead apply tensor-train decomposition to embeddings and low-rank approximation to dense weights, respectively. Recent advances have shifted toward dynamic, instance-aware model adaptation. DIET~\cite{fu2024DIET} and Gater~\cite{chen2019you} perform importance analysis to identify customized parameters. CHORD~\cite{liu2025chord} uses device-cloud collaboration to generate personalized channel-wise mixed-precision quantization strategies for on-device recommendation, among other device-cloud approaches~\cite{cocorrrec2025,lsc4rec2025}. On the input side, ATP-LLaVA dynamically removes redundant tokens~\cite{Ye_2025_CVPR}. 
PerSRec~\cite{zhang2025efficient} and CAUSE~\cite{liu2026accelerating} compress user histories into learnable or category-based tokens, respectively, but retain fixed compression budgets. RACE instead jointly adapts model and input size to instance complexity for more flexible resource allocation.

\subsection{Model Uncertainty in Recommendation}
Uncertainty quantification characterizes predictive reliability in recommender systems~\cite{10.1145/3584021}. Common point-wise estimators, namely Bayesian neural networks~\cite{blundell2015weight}, MC Dropout~\cite{gal2016dropout}, and deep ensembles~\cite{lakshminarayanan2017simple}, build predictive distributions from posterior weights, stochastic forward passes, and independently trained models, respectively, summarized by item-level variance or entropy. Pairwise-margin heuristics use score gaps to approximate local relative confidence~\cite{thurstone}. In recommendation, these signals mainly serve calibration and uncertainty-aware re-ranking~\cite{heusspredictive2023,yangmitigating2023}, and LiDu~\cite{li2025recommendersselfawarelabelfreerecommendation} targets label-free performance estimation. These works assess or refine predictions, whereas RACE anticipates list-wise uncertainty to guide capacity-adaptive inference.

\section{Method}
In this section, we first formalize our problem in Section~\ref{sec:problem}, then introduce an uncertainty-guided resource allocation mechanism in Section~\ref{sec:uncertainty}, followed by multi-granularity customization strategies for model structure (Section~\ref{sec:structure}) and input sequences (Section~\ref{sec:sequence}).

\subsection{Preliminary}
\label{sec:problem}
In efficient sequential recommendation scenario, we consider a system with an item set $\mathcal{I} = \{i_1, i_2, \ldots, i_{|\mathcal{I}|}\}$ and a user set $\mathcal{U} = \{u_1, u_2, \ldots, u_{|\mathcal{U}|}\}$. For each user $u \in \mathcal{U}$, their interaction history is represented as a chronologically ordered sequence $S^u = [s^u_1, \allowbreak s^u_2, \allowbreak \ldots, \allowbreak s^u_T]$, where $s^u_t \in \mathcal{I}$ represents an item interacted at timestamp $t$, and $T$ denotes the sequence length.

Our task is to predict the user's next interaction item $s^u_{T+1} \in \mathcal{I}$ from the historical sequence $S^u$ with the maxlen $L$. Given the resource-constrained nature of mobile devices, providing computationally efficient personalized recommendation becomes our critical concern. Our primary objective can be formalized as:
\begin{equation}
\begin{aligned}
\max_{\pi^u} \quad & \sum_{u \in \mathcal{U}} R(\pi^u, M)(S^u), \\
\text{s.t.} \quad & C \leq C_{max},
\end{aligned}
\label{eq:objective}
\end{equation}
where $\pi^u = \{\tau^u, \tau'^u, \mathbf{m}^u, k^u\}$ is the instance-specific customization policy that determines: threshold $\tau^u$, $\tau'^u$ for channel and sequence pruning intensity, channel masks $\mathbf{m}^u$ for structure customization, and sequence filterings $k^u$ for sequence customization. The function $R$ applies $\pi^u$ to base model $M$ for personalized predictions. The constraint $C \leq C_{max}$ enforces the computational budget.

\begin{figure*}[t]
    \centering
    \includegraphics[width=0.98\linewidth]{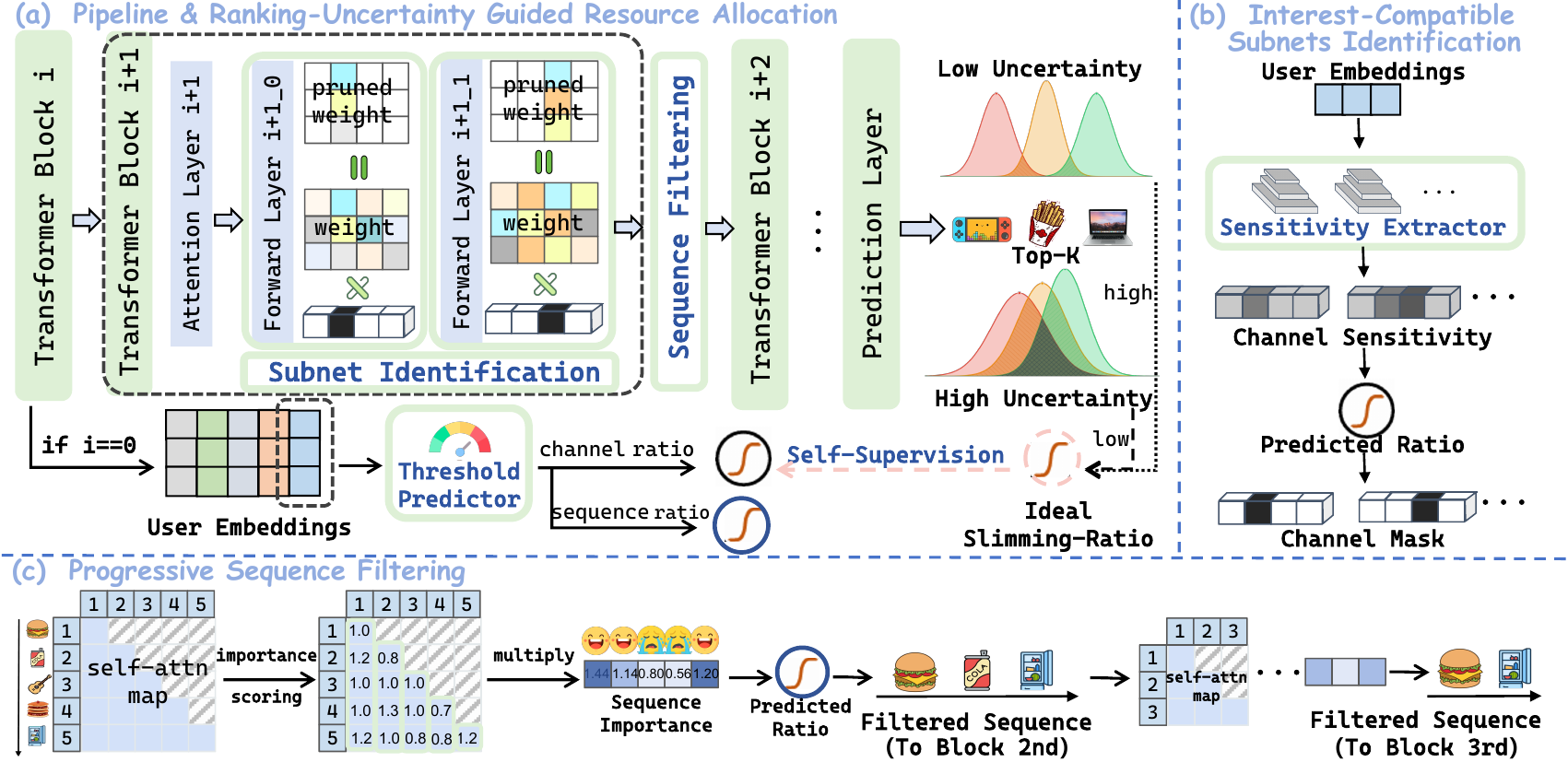}
    \vspace{-1mm}
    \caption{Overview of the RACE framework. Our approach consists of three key components: (a) \textbf{Ranking-Uncertainty Guided Resource Allocation}: RACE designs a list-wise ranking-uncertainty metric and transforms it into a proactive instance-complexity proxy to guide resource allocation, utilizing a self-supervised  threshold predictor to anticipate ratios early in inference.
    (b) \textbf{Interest-Compatible Subnets Identification}: Using the predicted ratio, a lightweight hypernetwork-style sensitivity extractor generates channel-wise masks to customize each feed-forward layer. (c) \textbf{Progressive Sequence Filtering}: Using the predicted ratio, RACE calculates relative importance scores to retain semantically relevant items while suppressing noise block by block. }
    \label{fig:main}
    \vspace{-1mm}
\end{figure*}

\subsection{Ranking-Uncertainty Guided Resource Allocation}
\label{sec:uncertainty}
 
The foundation of adaptive inference lies in measuring model uncertainty to serve as the proxy for instance complexity. We propose a list-wise ranking-uncertainty metric that captures the model's confidence in its predictions, which then guides our resource allocation strategy through a self-supervised learning mechanism.
 
\subsubsection{List-Wise Ranking Uncertainty}
 
Traditionally, model uncertainty is measured via point-wise estimation methods, such as Bayesian Neural Networks~\cite{blundell2015weight}, MC Dropout~\cite{gal2016dropout}, and Deep Ensembles~\cite{lakshminarayanan2017simple}, which calculate the variance of isolated prediction scores. However, dropout methods require multiple forward passes and ensemble approaches necessitate maintaining multiple models, both incurring inference latency and storage burdens prohibitive for real-time service. We therefore adopt a variational Bayesian layer~\cite{harrison2024variational} as the prediction head, which maintains weights as $W \sim \mathcal{N}(\mu_W, \Sigma_W)$ and yields both a predictive mean $\boldsymbol{\mu} \in \mathbb{R}^{|\mathcal{I}|}$ and prediction variance $\boldsymbol{\sigma}^2 \in \mathbb{R}^{|\mathcal{I}|}$ for all items in a single forward pass.
 
Nevertheless, point-wise estimation alone is insufficient, since sequential recommendation is fundamentally a ranking problem where the relative order of items matters more than absolute scores.
We therefore lift uncertainty from individual scores to the \textbf{stability of the predicted Top-$K$ ordering}: rather than measuring how uncertain any single score is, we measure how stable the entire ordering remains under the predictive distribution. Concretely, we treat each pairwise comparison within the Top-$K$ as a probabilistic event and define the overall ranking uncertainty as the aggregated negative log-likelihood of these events, so that ambiguous orderings contribute more uncertainty than well-separated ones.

This construction reduces to deriving the pairwise outranking probability $p_{i \succ j}$ under the model's own predictive distribution, which the variational Bayesian layer makes available in closed form. We adopt a diagonal parameterization of $\Sigma_W$ across output dimensions~\cite{harrison2024variational}, so that the predictive components for different items are conditionally independent given the input. The linear projection $z = Wx$ then produces jointly Gaussian logits with class-wise independent components: $z_i \sim \mathcal{N}(\mu_i, \sigma_i^2)$, and the pairwise difference $z_i - z_j \sim \mathcal{N}(\mu_i - \mu_j,\, \sigma_i^2 + \sigma_j^2)$.
Applying Thurstone's law of comparative judgment~\cite{thurstone} to this Gaussian difference, the outranking probability admits the closed-form expression: 
\vspace{-1mm}
\begin{equation} 
p_{i \succ j} = P(z_i > z_j) = \Phi\!\left(\frac{\mu_i - \mu_j}{\sqrt{\sigma_i^2 + \sigma_j^2}}\right),
\label{eq:pairwise}
\end{equation}
where $\Phi(\cdot)$ is the standard normal CDF. Eq.~(\ref{eq:pairwise}) is thus a direct consequence of the model's own predictive distribution.
 
Let $\mathcal{K} = \{k_1, \ldots, k_K\}$ denote the indices of the top-$K$ items according to $\boldsymbol{\mu}$. The ranking uncertainty is then defined as the negative log-likelihood of all pairwise comparisons within the top-$K$ :
\vspace{-1mm}
\begin{equation}
\mathcal{U}_{\text{rank}} = -\sum_{i=1}^{K-1} \sum_{j=i+1}^{K} \log p_{k_i \succ k_j}.
\label{eq:ranking_uncertainty}
\end{equation}
We set $K=20$ in our experiments, intentionally larger than the recommendation evaluation cut-off  so that $\mathcal{U}_{\text{rank}}$ captures the broader structural stability of the predicted ranking rather than only the very top items.
Intuitively, $\mathcal{U}_{\text{rank}}$ approaches zero when the model is confident in its ordering, namely when large score gaps and small variances drive each $p_{i \succ j}$ toward $1$, and inflates whenever score gaps narrow or variances widen, signaling an ambiguous Top-$K$ ordering and a more complex instance.

\subsubsection{Theoretical Motivation}
\label{sec:theory}
We now provide the theoretical motivation underlying $\mathcal{U}_{\text{rank}}$. Two questions need to be addressed: first, why model uncertainty can serve as a signal of instance complexity; second, whether our list-wise construction preserves the probabilistic semantics that makes this signal meaningful.

\textit{Calibration makes uncertainty a principled complexity signal.} The use of predictive uncertainty as a signal of instance complexity is well-established in machine learning, with roots in curriculum learning~\cite{bengio2009curriculum} and uncertainty quantification~\cite{hullermeier2021aleatoric}. This usage rests on calibration~\cite{guo2017calibration,jeon2024calibration}: when a model's confidence aligns with its empirical accuracy, higher predictive uncertainty signals harder instances. To ground $\mathcal{U}_{\text{rank}}$ on a calibrated source of uncertainty, we adopt the variational Bayesian last layer (VBLL)~\cite{harrison2024variational} as the prediction head. Trained with a variational evidence lower bound, VBLL achieves lower expected calibration error than deterministic baselines across regression and classification tasks~\cite{harrison2024variational}. Its predictive variance $\sigma^2$ quantifies epistemic uncertainty~\cite{kendall2017uncertainties}: a larger value reflects insufficient evidence for a confident prediction and therefore serves as a proxy for instance complexity.

\textit{The list-wise extension preserves this probabilistic grounding.} The calibration argument above concerns the predictive distribution over a single outcome, while sequential recommendation operates on ordered lists. $\mathcal{U}_{\text{rank}}$ bridges this gap through two probabilistic constructions. The pairwise outranking probability $p_{i \succ j}$ (Eq.~\ref{eq:pairwise}) is the closed-form solution of the Bayesian Thurstone model~\cite{thurstone} under Gaussian distribution. 
Aggregating these probabilities through the negative log-likelihood (Eq.~\ref{eq:ranking_uncertainty}) induces a coherent measure of list-wise ranking confidence, approaching its minimum when all top-K pairs are confidently separated. The resulting $\mathcal{U}_{\text{rank}}$ is therefore a functional of the predictive distribution~\cite{kendall2017uncertainties}, reflecting the structural stability of the predicted ordering as a list-wise complexity signal. We empirically verify this argument in Section~\ref{sec:uncertainty_validation}.

\subsubsection{Self-Supervised Threshold Prediction Training}

While ranking uncertainty provides a reliable measure of instance complexity, computing it requires a full forward pass including the final prediction layer. To enable intelligent resource allocation, we need to anticipate the ranking uncertainty early in inference. 

Specifically, we design a \textbf{lightweight threshold predictor} $g_{\phi}: \mathbb{R}^d \to [0, 1]$ to predict the ideal slimming rate.
The predictor takes a context vector $\mathbf{c} \in \mathbb{R}^d$ as a user embedding, which we obtain by extracting the last valid token representation from early-layer representations (details in Section~\ref{sec:structure}). The threshold predictor is implemented with a simple two-layer MLP: 
\begin{equation}
\tau = g_{\phi}(\mathbf{c}) ,
\end{equation}
where $\tau \in [0, 1]$ directly controls structure customization, while a linearly transformed variant $\tau'$ governs sequence customization (detailed in Sections~\ref{sec:structure} and~\ref{sec:sequence}).

To early capture uncertainty, we establish a \textbf{self-supervised learning} objective. We compute the ground-truth ranking uncertainty $\mathcal{U}_{\text{rank}}$ and normalize it using  the current mini-batch statistics:
\begin{equation}
\tilde{\mathcal{U}} = \frac{\mathcal{U}_{\text{rank}} - \mathbb{E}[\mathcal{U}_{\text{rank}}]}{\text{std}(\mathcal{U}_{\text{rank}}) + \epsilon}.
\end{equation}
We then map the normalized uncertainty to a target threshold $\tau^*$ as our ideal slimming rate through a sigmoid transformation:
\begin{equation}
\tau^* = \sigma\left(-\tilde{\mathcal{U}} + b_{\text{center}}\right),
\end{equation}
The uncertainty-derived target $\tau^*$ is detached during predictor training.
where $b_{\text{center}} = \log\left(\frac{r_0}{1-r_0}\right)$ is designed to center the threshold distribution around a target pruning ratio $r_0$. The negative sign ensures that higher uncertainty (difficult instances) leads to lower ratios, which results in less aggressive pruning and greater model capacity. During training, the threshold predictor is guided by the pseudo-label $\tau^*$ using the mean squared error:
\begin{equation}
\mathcal{L}_{\text{threshold}} = \mathbb{E}_{S \sim \mathcal{D}}\left[(\tau - \tau^*)^2\right].
\label{eq:threshold_loss}
\end{equation}

This self-supervised approach allows the predictor to learn the mapping from early representations to appropriate resource allocation without requiring external labels or manual annotations.

\subsection{Interest-Compatible Subnet Identification}
\label{sec:structure}

Once the instance-specific resource budget (compression ratio $\tau$) is determined by the uncertainty predictor, the next challenge is deciding \textit{which} specific parameters to retain. 
Existing methods often rely on external auxiliary models to extract user features~\cite{fu2024DIET}, which ironically introduces additional computational overhead. 
To address this, RACE adopts an \textbf{Intrinsic Feature Reuse} strategy. We posit that the model's intermediate representations already contain useful information about user intent to guide its own structural adaptation, eliminating the need for heavy external extractors.

Formally, let $\mathbf{H}^{(l)} \in \mathbb{R}^{T \times d}$ denote the hidden states at the $l$-th Transformer layer. We extract the representation of the last valid item in the sequence as the dynamic user embedding $\mathbf{c}^{(l)}$, which serves as the context for decision-making. 
To determine the importance of each channel in the Feed-Forward Network (FFN), we employ a lightweight \textbf{Sensitivity Extractor}. This module, acting as a hypernetwork-style scorer, projects the user embedding $\mathbf{c}^{(l)}$ into channel-wise sensitivity scores $\mathbf{s}^{(l)} \in \mathbb{R}^{d_{ff}}$:
\begin{equation}
    \mathbf{s}^{(l)} = \sigma(\mathbf{W}_s \mathbf{c}^{(l)} + \mathbf{b}_s),
\end{equation}
where $\mathbf{W}_s$ and $\mathbf{b}_s$ are learnable parameters, and $\sigma$ is the sigmoid function. 
These scores represent the potential contribution of each neuron to the current user's inference.

To construct the channel mask $\mathbf{m}^{(l)}$,  we utilize the predicted channel pruning ratio $\tau$ to perform dynamic truncation. 
Specifically, we identify the threshold value $\delta$ at the $\tau$-quantile of $\mathbf{s}^{(l)}$ and generate the mask as follows:
\begin{equation}
    \mathbf{m}^{(l)}_i = \mathbb{I}(\mathbf{s}^{(l)}_i \geq \delta),
\end{equation}
where $\mathbb{I}(\cdot)$ is the indicator function. Thus, $\tau$ is the pruned-channel fraction, and the default $\tau=0.8$ prunes approximately 80\% of the channels. 
During training, we use a Straight-Through Estimator (STE): the hard mask is detached, while gradients are passed through the continuous channel scores. 
By applying this mask with the FFN weights, RACE dynamically carves out \textit{interest-compatible subnets}. 
\subsection{Progressive Sequence Filtering}
\label{sec:sequence}

Guided by the predicted ratio $\tau'$, RACE  implements a dynamic data-slimming strategy to identify crucial items. Drawing inspiration from token pruning in natural language processing~\cite{kim2022learned,goyal2020power,ye2021tr}, we propose \textbf{Progressive Sequence Filtering} to eliminate redundancy of user interaction history block by block, ensuring the computational budget is concentrated on the most informative items.

In Transformer-based architectures, the self-attention map $\mathbf{A}^{(l)} \in \mathbb{R}^{H \times T \times T}$ naturally encodes the contribution of each item. However, in sequential recommendation, standard attention scores are subject to causal masking and can be volatile across layers. To capture the relative significance for each interaction, we design a \textit{cumulative importance metric} based on the head-averaged attention map $\bar{\mathbf{A}}^{(l)} = \frac{1}{H} \sum \mathbf{A}^{(l)}$. As illustrated in Figure~\ref{fig:intro}(c), for each item $t \in \{1, \dots, T\}$, its importance score $s_t$ is formulated as the normalized product of its weighted contributions across the sequence:
\begin{equation}
    s_t = \frac{\tilde{s}_t}{\sum_{j=1}^{T} \tilde{s}_j + \epsilon}, \quad \text{where} \quad \tilde{s}_t = \prod_{j=t}^{T} \left( \bar{\mathbf{A}}^{(l)}_{j,t} \cdot \eta(j) \right).
\end{equation}
Here, $\eta(j) = j$ serves as a position-aware factor that linearly scales attention weights by the query position $j$, effectively prioritizing later queries that encode more recent user interests. The product operator ($\prod$) functions as a cumulative filter. This approach accumulates attention scores from later, more informative queries, effectively prioritizing items that consistently contribute to the user's current interests.

Using the uncertainty-driven layer-wise pruning ratio $\tau'$, block $l$ retains $n_l^{\mathrm{keep}}=\lceil |S^{(l)}|(1-\tau')\rceil$ items:
\begin{equation}
    S^{(l+1)} = \operatorname{TopK}\!\left(S^{(l)}, \mathbf{s}, n_l^{\mathrm{keep}}\right).
\end{equation}
In the \textbf{training} phase, we employ a "soft" filtering approach by masking unselected tokens to maintain batch stability while simulating the pruning effect. By contrast, during \textbf{inference}, unselected tokens are dropped to achieve immediate FLOPs reduction. 

\begin{table*}[!t]
  \caption{Overall performance comparison of RACE and competitive baselines across four datasets and two representative backbones. For metrics $NDCG@K$ and $Hit@K$, $K=5$ is used by default. $\uparrow$ and $\downarrow$ denote that larger and smaller values are preferred, respectively. \textbf{Bold} font indicates the best performance in each column.}
\vspace{-1mm}
\label{tab:experiment_results}
\centering
\resizebox{1.0\textwidth}{!}{
    \begin{tabular}{@{}cl*{12}{c}@{}}
    \toprule
     \multirow{2}{*}{\textbf{Model}} & \multirow{2}{*}{\textbf{Method}}
     & \multicolumn{3}{c}{\textbf{MovieLens-100k}}
     & \multicolumn{3}{c}{\textbf{Douban}}
     & \multicolumn{3}{c}{\textbf{Amazon-CD}}
     & \multicolumn{3}{c}{\textbf{Amazon-Game}} \\
     \cmidrule(lr){3-5}\cmidrule(lr){6-8}\cmidrule(lr){9-11}\cmidrule(lr){12-14}
     & & \textbf{NDCG $\uparrow$} & \textbf{Hit $\uparrow$} & \textbf{FLOPs $\downarrow$}
     & \textbf{NDCG $\uparrow$} & \textbf{Hit $\uparrow$} & \textbf{FLOPs $\downarrow$}
     & \textbf{NDCG $\uparrow$} & \textbf{Hit $\uparrow$} & \textbf{FLOPs $\downarrow$}
     & \textbf{NDCG $\uparrow$} & \textbf{Hit $\uparrow$} & \textbf{FLOPs $\downarrow$} \\
     \midrule

     \multirow{11}{*}{\textbf{SASRec}} & \texttt{Normal} & 0.0299 & 0.0471 & 1.8432 & 0.0141 & 0.0193 & 1.8432 & 0.0250 & 0.0295 & 1.8432 & 0.0334 & 0.0418 & 1.8432 \\
     & \texttt{Finetune} & 0.0268 & 0.0442 & 1.8432 & 0.0131 & 0.0174 & 1.8432 & 0.0251 & 0.0297 & 1.8432 & 0.0338 & 0.0424 & 1.8432 \\
     & \texttt{SNIP} & 0.0285 & 0.0464 & 0.8352 & 0.0109 & 0.0153 & 0.8352 & 0.0240 & 0.0280 & 0.8352 & 0.0331 & 0.0401 & 0.8352 \\
     & \texttt{GEM} & 0.0330 & 0.0516 & 0.8352 & 0.0144 & 0.0199 & 0.8352 & 0.0263 & 0.0312 & 0.8352 & 0.0343 & 0.0422 & 0.8352 \\
     & \texttt{APG} & 0.0338 & 0.0536 & 0.8947 & 0.0224 & 0.0305 & 0.8947 & 0.0267 & 0.0312 & 0.8947 & 0.0385 & 0.0468 & 0.8947 \\
     & \texttt{DIET} & 0.0328 & 0.0521 & 1.0163 & 0.0083 & 0.0113 & 0.8167 & 0.0254 & 0.0287 & 0.9010 & 0.0397 & 0.0481 & 0.8925 \\
     & \texttt{ATP-LLAVA} & 0.0240 & 0.0377 & 0.9010 & 0.0086 & 0.0134 & 0.9491 & 0.0254 & 0.0298 & 1.0143 & 0.0311 & 0.0373 & 0.9815 \\
     & \texttt{PerSRec} & 0.0259 & 0.0411 & 0.8704 & 0.0067 & 0.0103 & 0.8704 & 0.0106 & 0.0150 & 0.8704 & 0.0175 & 0.0256 & 0.8704 \\
     & \texttt{CAUSE} & 0.0277 & 0.0442 & 0.8704 & 0.0047 & 0.0070 & 0.8704 & 0.0248 & 0.0291 & 0.8704 & 0.0316 & 0.0403 & 0.8704 \\
     \rowcolor{lightblue}
     \cellcolor{white} & \texttt{RACE} & \textbf{0.0427} & \textbf{0.0646} & \textbf{0.8025} & \textbf{0.0261} & \textbf{0.0337} & \textbf{0.7928} & \textbf{0.0288} & \textbf{0.0329} & \textbf{0.7549} & \textbf{0.0402} & \textbf{0.0490} & \textbf{0.7660} \\
     \rowcolor{lightgray!40}
     \cellcolor{white} & \texttt{Improv.} & 42.81\% & 37.15\% & 2.30$\times$ & 85.11\% & 74.61\% & 2.32$\times$ & 15.20\% & 11.50\% & 2.44$\times$ & 20.36\% & 17.22\% & 2.41$\times$ \\

     \midrule

     \multirow{11}{*}{\textbf{BSARec}} & \texttt{Normal} & 0.0256 & 0.0409 & 4.1370 & 0.0206 & 0.0275 & 4.1370 & 0.0368 & 0.0431 & 4.1370 & 0.0325 & 0.0404 & 4.1370 \\
     & \texttt{Finetune} & 0.0305 & 0.0475 & 4.1370 & 0.0164 & 0.0216 & 4.1370 & 0.0383 & 0.0451 & 4.1370 & 0.0312 & 0.0395 & 4.1370 \\
     & \texttt{SNIP} & 0.0322 & 0.0509 & 1.7454 & 0.0305 & 0.0391 & 1.7454 & 0.0418 & 0.0492 & 1.7454 & 0.0335 & 0.0429 & 1.7454 \\
     & \texttt{GEM} & 0.0298 & 0.0465 & 1.7454 & 0.0219 & 0.0302 & 1.7454 & 0.0388 & 0.0455 & 1.7454 & 0.0307 & 0.0395 & 1.7454 \\
     & \texttt{APG} & 0.0358 & 0.0564 & 2.1665 & 0.0282 & 0.0369 & 2.1665 & 0.0409 & 0.0481 & 2.1665 & 0.0327 & 0.0420 & 2.1665 \\
     & \texttt{DIET} & 0.0346 & 0.0540 & 2.3845 & 0.0203 & 0.0269 & 2.3490 & 0.0419 & 0.0482 & 2.4660 & 0.0339 & 0.0424 & 1.8703 \\
     & \texttt{ATP-LLAVA} & 0.0287 & 0.0470 & 2.8243 & 0.0111 & 0.0163 & 2.6770 & 0.0083 & 0.0105 & 2.5161 & 0.0108 & 0.0156 & 2.2927 \\
     & \texttt{PerSRec} & 0.0279 & 0.0442 & 2.0173 & 0.0156 & 0.0216 & 2.0173 & 0.0349 & 0.0396 & 2.0173 & 0.0328 & 0.0395 & 2.0173 \\
     & \texttt{CAUSE} & 0.0266 & 0.0421 & 2.0173 & 0.0135 & 0.0192 & 2.0173 & 0.0381 & 0.0438 & 2.0173 & 0.0318 & 0.0395 & 2.0173 \\
     \rowcolor{lightblue}
     \cellcolor{white} & \texttt{RACE} & \textbf{0.0396} & \textbf{0.0600} & \textbf{1.7148} & \textbf{0.0327} & \textbf{0.0418} & \textbf{1.6473} & \textbf{0.0482} & \textbf{0.0542} & \textbf{1.6432} & \textbf{0.0476} & \textbf{0.0576} & \textbf{1.6424} \\
     \rowcolor{lightgray!40}
     \cellcolor{white} & \texttt{Improv.} & 54.69\% & 46.70\% & 2.41$\times$ & 58.74\% & 52.00\% & 2.51$\times$ & 31.01\% & 25.82\% & 2.52$\times$ & 46.46\% & 42.57\% & 2.52$\times$ \\

    \bottomrule
    \end{tabular}

}
\end{table*}

\subsection{Loss Function}
\label{sec:loss}
Our training objective jointly optimizes recommendation accuracy, resource efficiency (Sections~\ref{sec:structure} and~\ref{sec:sequence}), and uncertainty-aware adaptation (Section~\ref{sec:uncertainty}). The complete loss function consists of three components:
\begin{equation}
\mathcal{L}_{\text{total}} = \mathcal{L}_{\text{rec}} + \lambda_1 \mathcal{L}_{\text{sparsity}} + \lambda_2 \mathcal{L}_{\text{threshold}}.
\label{eq:total_loss}
\end{equation}
where
$\mathcal{L}_{\text{rec}}$ is the recommendation loss, combining cross-entropy and KL divergence for Bayesian inference. $\mathcal{L}_{\text{sparsity}} = (\rho - r_{\text{target}})^2$ penalizes deviation from average sparsity ratio $r_{\text{target}}$, where $\rho$ is the average channel slimming ratio. $\mathcal{L}_{\text{threshold}}$ (Eq.~\ref{eq:threshold_loss}) trains the threshold predictor to anticipate the ideal pruning ratio.  $\lambda_1$ and $\lambda_2$ are scaling coefficients used to balance the trade-off between accuracy, average sparsity, and dynamic resource allocation.

\section{EXPERIMENTS}

We conduct extensive experiments on four public datasets and
answer the following research questions:

\begin{itemize}[leftmargin=*,nosep]
    \item \textbf{RQ1:} Does RACE outperform existing methods in 
   accuracy while simultaneously reducing computational
    cost?
    
    \item \textbf{RQ2:} Is each component of RACE necessary and 
    superior to existing alternatives?

    \item \textbf{RQ3:} Does list-wise ranking-uncertainty reliably reflect instance
    complexity and effectively drive RACE's adaptive allocation?

    \item \textbf{RQ4:} Does RACE achieve stable and robust training 
    behavior?

    \item \textbf{RQ5:} Does RACE achieve efficient, adaptive inference under varying resource constraints?

\end{itemize}

\subsection{Experimental Setup}

\subsubsection{Datasets}
The experiments are conducted on four datasets:  MovieLens-100k\footnote{https://grouplens.org/datasets/movielens/100k}, Douban~\cite{DTCDR}, Amazon-CD\footnote{https://nijianmo.github.io/amazon/index.html} and Amazon-Game.
Following the 10-core setting, we exclude users and items with fewer than 10 interactions. For each user, the most recent 10\% of interactions form the test set and the rest are used for training. 

\subsubsection{Baselines}
We adopt SASRec~\cite{kang2018self} and BSARec~\cite{shin2024attentive} as backbones and compare RACE with nine baselines: (1) \textbf{Full-precision Methods}: Normal and Finetune. (2) \textbf{Compression-based Methods}: 
\begin{itemize}[leftmargin=*,nosep]
    \item \textbf{SNIP}~\cite{lee2018snip}: Prunes connections by saliency before training.
    \item \textbf{GEM}~\cite{sreenivasan2022rare}: Finds high-performance subnets at initialization.
    \item \textbf{APG}~\cite{yan2022apg}: Decomposes weights via low-rank approximation.
    \item \textbf{DIET}~\cite{fu2024DIET}: Customizes model parameters for individual devices through multi-level sensitivity analysis.
    \item \textbf{ATP-LLAVA}~\cite{Ye_2025_CVPR}: Prunes redundant tokens dynamically.
    \item \textbf{PerSRec}~\cite{zhang2025efficient}: Compresses histories into a fixed number of personalized learnable tokens, keeping recent interactions.
    \item \textbf{CAUSE}~\cite{liu2026accelerating}: Merges histories into category-based tokens.
\end{itemize}

\subsubsection{Evaluation Metrics}
We evaluate recommendation quality with Hit@5 (Hit) and NDCG@5 (NDCG) under full ranking over the entire item set to avoid candidate-sampling bias, and measure computational cost by backbone inference FLOPs (MFLOPs).

\subsubsection{Implementation Detail}
For all datasets, we set the maximum sequence length to 20 with zero padding, the item embedding dimension to 64, and four Transformer blocks for both backbones. The loss coefficients $\lambda_1$ and $\lambda_2$ (Section~\ref{sec:loss}) are set to 1 and 0.5, respectively. For compression-based methods, we tune their hyperparameters to reach sparsity similar to ours without much degradation.

\subsection{Overall Performance(RQ1)}
\label{sec:overall_performance}

Table~\ref{tab:experiment_results} presents the overall performance comparison across four datasets and two backbones. From the results, we can observe that  \textbf{Obs.1:  RACE consistently achieves the best performance in both accuracy and efficiency across all settings.} Averaged over the four datasets, RACE improves NDCG and Hit by 40.87\% and 35.12\% on SASRec while reducing FLOPs by 2.37$\times$. On BSARec, the corresponding average improvements are 47.73\% and 41.77\%, together with a 2.49$\times$ FLOPs reduction.

Among the baselines, finetuning with limited data does not guarantee performance improvement. SNIP and GEM apply uniform masks across all instances and fail to adapt to instance interests and complexity. APG and DIET customize the model side through low-rank decomposition and sensitivity analysis, respectively, and ATP-LLAVA dynamically prunes tokens, yet these single-side methods show unstable accuracy across datasets. PerSRec and CAUSE compress user histories into learnable or category-based tokens but retain fixed compression budgets. RACE also improves over the strongest competing baseline in representative settings: on Douban with SASRec, it achieves 0.0261 NDCG, 16.52\% above APG (0.0224); on Amazon-CD with BSARec, it achieves 0.0482 NDCG, 15.04\% above DIET (0.0419), while using fewer FLOPs than CAUSE (1.6432 vs. 2.0173 MFLOPs).

\subsection{Ablation Study (RQ2)}
\label{sec:ablation}
 
Table~\ref{tab:ablated} reports the ablation results, and Tables~\ref{tab:uncertainty_compact} and~\ref{tab:uncertainty_ablation} compare two key components of RACE with alternative designs.

\begin{table}[t]
\vspace{1.5mm}
\caption{Ablation experiment.}
\vspace{-1mm}
\label{tab:ablated}
\centering
\scriptsize
\renewcommand{\arraystretch}{0.93}
\resizebox{0.94\linewidth}{!}{
    \begin{tabular}{@{}clcccc@{}}
    \toprule
     \multirow{2}{*}{\textbf{Model}} & \multirow{2}{*}{\textbf{Ablated Model}}
     & \multicolumn{2}{c}{\textbf{MovieLens-100k}} & \multicolumn{2}{c}{\textbf{Douban}} \\
     \cmidrule(lr){3-4}\cmidrule(lr){5-6}
     & & \textbf{NDCG $\uparrow$} & \textbf{Hit $\uparrow$} & \textbf{NDCG $\uparrow$} & \textbf{Hit $\uparrow$} \\
     \midrule
     \multirow{6}{*}{\textbf{SASRec}} & \texttt{w/o uncertainty} & 0.0299 & 0.0471 & 0.0141 & 0.0193 \\
     & \texttt{w/o customization} & 0.0263 & 0.0447 & 0.0139 & 0.0193 \\
     & \texttt{w/o sequence filtering} & 0.0388 & 0.0592 & 0.0229 & 0.0296 \\
     & \texttt{w/o subnet identification} & 0.0296 & 0.0492 & 0.0104 & 0.0171 \\
     & \texttt{w/o dynamic ratio} & 0.0376 & 0.0584 & 0.0250 & 0.0321 \\
     \rowcolor{lightblue}
     \cellcolor{white} & \textbf{\texttt{RACE}} & \textbf{0.0427} & \textbf{0.0646} & \textbf{0.0261} & \textbf{0.0337} \\
     \midrule
     \multirow{6}{*}{\textbf{BSARec}} & \texttt{w/o uncertainty} & 0.0256 & 0.0409 & 0.0206 & 0.0275 \\
     & \texttt{w/o customization} & 0.0275 & 0.0441 & 0.0095 & 0.0150 \\
     & \texttt{w/o sequence filtering} & 0.0310 & 0.0507 & 0.0229 & 0.0304 \\
     & \texttt{w/o subnet identification} & 0.0268 & 0.0443 & 0.0094 & 0.0141 \\
     & \texttt{w/o dynamic ratio} & 0.0326 & 0.0520 & 0.0233 & 0.0315 \\
     \rowcolor{lightblue}
     \cellcolor{white} & \textbf{\texttt{RACE}} & \textbf{0.0396} & \textbf{0.0600} & \textbf{0.0327} & \textbf{0.0418} \\
    \bottomrule
    \end{tabular}
}
\end{table}

\vspace{1mm}
\subsubsection{Component Ablation}

\textbf{Obs.2: Every component of RACE plays a meaningful and complementary role.} 

\begin{itemize}[leftmargin=*,nosep]
    \item \textit{w/o Uncertainty}  collapses RACE to the unified backbone and eliminates all gains.
    \item \textit{w/o Customization} replaces the backbone's prediction layer with a variational Bayesian layer for uncertainty estimation, showing that uncertainty alone cannot model user interests accurately and even degrades performance.
    \item \textit{w/o Sequence Filtering} keeps only interest-aware subnet identification, showing that mitigating model redundancy contributes meaningfully, with a 29.8\% NDCG improvement on MovieLens-100K with SASRec.
    \item \textit{w/o Subnet Identification} retains sequence filtering but disables channel-wise subnet customization, showing that reducing data redundancy alone sometimes degrades performance, e.g., on Douban.
    \item \textit{w/o Dynamic Ratio} applies both customizations with a fixed pruning ratio, showing that jointly alleviating model and data redundancy benefits both accuracy and efficiency and outperforms the full-precision model.
    \item \textit{RACE} further allocates model and input size dynamically under uncertainty guidance. Its gap over w/o Dynamic Ratio (e.g., over 40\% on Douban/BSARec) validates dynamic resource allocation based on instance complexity.
\end{itemize}

\subsubsection{Comparison with Alternative Designs}

\textbf{Obs.3: RACE's design choices outperform alternative designs in both uncertainty estimation and sequence filtering.} 

\begin{table}[t]
\vspace{1.5mm}
\caption{Comparison of alternative uncertainty designs for guiding dynamic thresholds on Douban.}
\vspace{-1mm}
\label{tab:uncertainty_compact}
\centering
\scriptsize
\renewcommand{\arraystretch}{0.93}
\resizebox{0.94\linewidth}{!}{
    \begin{tabular}{@{}lcccc@{}}
    \toprule
    \multirow{2}{*}{\textbf{Uncertainty Design}}
        & \multicolumn{2}{c}{\textbf{SASRec}}
        & \multicolumn{2}{c}{\textbf{BSARec}} \\
    \cmidrule(lr){2-3}\cmidrule(lr){4-5}
        & \textbf{NDCG} $\uparrow$
        & \textbf{Hit} $\uparrow$
        & \textbf{NDCG} $\uparrow$
        & \textbf{Hit} $\uparrow$ \\
    \midrule
    MC Dropout      & 0.0216 & 0.0292          & 0.0284 & 0.0366 \\
    Ensemble        & 0.0197 & 0.0266          & 0.0261 & 0.0344 \\
    Mean Variance   & 0.0248 & 0.0317          & 0.0310 & 0.0415 \\
    Entropy         & 0.0227 & 0.0311          & 0.0296 & 0.0389 \\
    Pairwise Margin & 0.0228 & 0.0289          & 0.0323 & 0.0407 \\
    LIDU            & 0.0259 & \textbf{0.0337} & 0.0289 & 0.0378 \\
    \midrule
    \textbf{RACE}  & \textbf{0.0261} & \textbf{0.0337}
                    & \textbf{0.0327} & \textbf{0.0418} \\
    \bottomrule
    \end{tabular}
}
\end{table}

\textbf{\textit{Comparison with alternative uncertainty designs.}}
Table~\ref{tab:uncertainty_compact} compares two levels of uncertainty design: predictive uncertainty estimation and ranking-uncertainty formulation. At the first level, we replace the variational Bayesian last layer (VBLL)~\cite{harrison2024variational} in RACE's prediction head with MC Dropout~\cite{gal2016dropout} and Ensemble~\cite{lakshminarayanan2017simple}, both restricted to the prediction layer for a compute-aligned comparison. VBLL produces predictive means and variances in one pass without repeated stochastic passes or multiple heads, and yields NDCG@5 of 0.0261/0.0327 versus 0.0216/0.0284 for the stronger alternative on SASRec/BSARec. At the second level, with VBLL fixed, Mean Variance and Entropy provide point-wise summaries~\cite{10.1145/3584021}, and Pairwise Margin and LIDU~\cite{li2025recommendersselfawarelabelfreerecommendation} add local ranking information, whereas our list-wise ranking uncertainty models the distributional stability of the entire top-$K$ list by considering interdependencies among candidates. RACE achieves the best accuracy on both backbones, supporting VBLL as an efficient technical choice and, more importantly, validating our list-wise uncertainty design for dynamic threshold prediction.

\begin{table}[t]
\caption{Comparison of different sequence-filtering methods on Douban and Amazon-Game datasets.}
\vspace{-1mm}
\label{tab:uncertainty_ablation}
\centering
\scriptsize
\renewcommand{\arraystretch}{0.93}
\resizebox{0.94\linewidth}{!}{
    \begin{tabular}{@{}clcccc@{}}
    \toprule
    \multirow{2}{*}{\textbf{Model}} & \multirow{2}{*}{\textbf{Method}} & \multicolumn{2}{c}{\textbf{Douban}} & \multicolumn{2}{c}{\textbf{Amazon-Game}} \\
    \cmidrule(lr){3-4}\cmidrule(lr){5-6}
     & & \textbf{NDCG} $\uparrow$ & \textbf{Hit} $\uparrow$ & \textbf{NDCG} $\uparrow$ & \textbf{Hit} $\uparrow$ \\
    \midrule
    \multirow{3}{*}{\textbf{SASRec}}
    & Mean & 0.0238 & 0.0312 & 0.0393 & 0.0471 \\
    & Masked Mean & 0.0241 & 0.0318 & 0.0390 & 0.0474 \\
    & \textbf{RACE} & \textbf{0.0261} & \textbf{0.0337} & \textbf{0.0402} & \textbf{0.0490} \\
    \midrule
    \multirow{3}{*}{\textbf{BSARec}}
    & Mean & 0.0310 & 0.0395 & 0.0472 & 0.0567 \\
    & Masked Mean & 0.0307 & 0.0397 & \textbf{0.0477} & 0.0568 \\
    & \textbf{RACE} & \textbf{0.0327} & \textbf{0.0418} & 0.0476 & \textbf{0.0576} \\
    \bottomrule
    \end{tabular}
}
\end{table}
 
\textbf{\textit{Comparison with other item-importance metrics.}} Beyond uncertainty estimation, Table~\ref{tab:uncertainty_ablation} compares item-importance metrics for sequence filtering. The global average of attention scores across heads (Mean)~\cite{Ye_2025_CVPR}, common in prior work, ignores the causal structure of sequential recommendation and causes an importance mismatch. Applying a causal mask to the average (Masked Mean) still suffers from positional bias, where earlier items receive disproportionate attention regardless of their relevance. RACE's cumulative and position-aware metric prioritizes recent and semantically core interactions, achieving the best performance on seven of the eight reported metrics and remaining comparable on the other. For instance, on Douban with SASRec, RACE raises NDCG to 0.0261, compared with 0.0238 for Mean and 0.0241 for Masked Mean.

\subsection{Uncertainty-Driven Resource Allocation Analysis (RQ3)}
\label{sec:uncertainty_validation}

\begin{figure}[t]
    \centering
    \vspace{1mm}
     \includegraphics[width=0.96\linewidth]{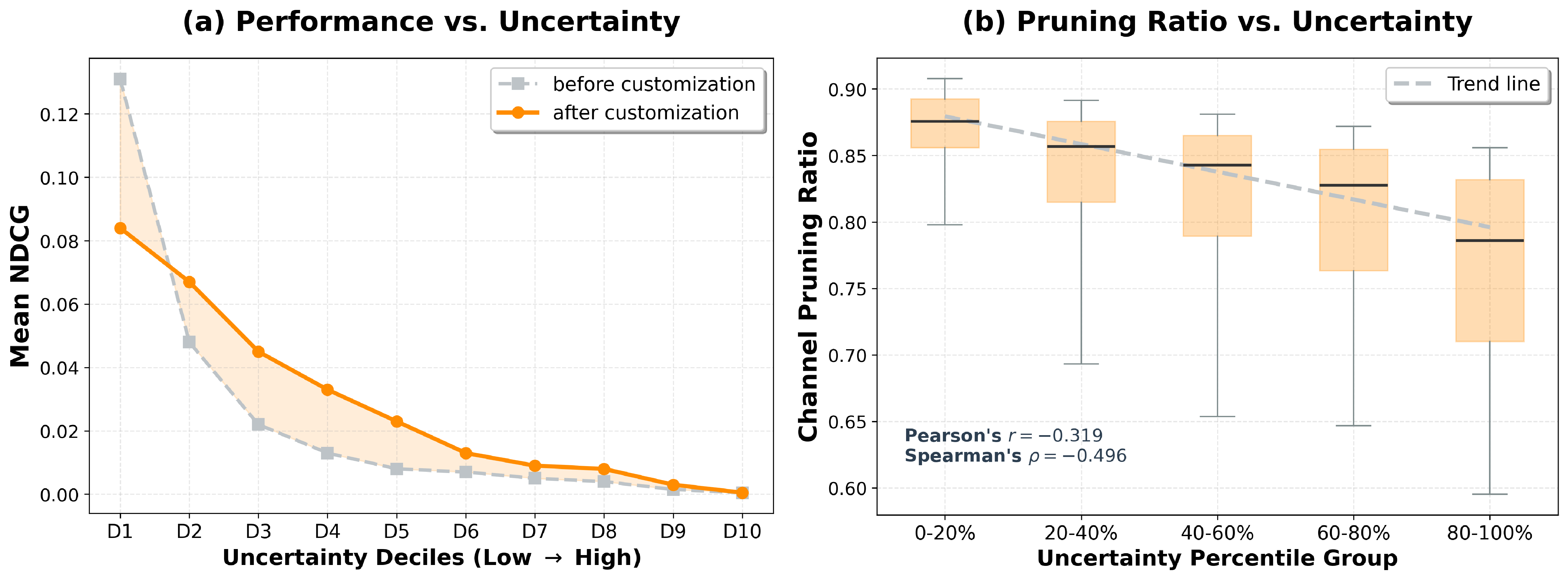}
    \vspace{-2mm}
    \caption{Uncertainty Analysis: (a) Performance (before or after customization) vs. Uncertainty (b) Predicted Channel Pruning Ratio vs. Uncertainty. }
    \vspace{-1mm}
    \label{fig:relation}
\end{figure}

Figure~\ref{fig:relation} examines ranking uncertainty from two perspectives: its correlation with recommendation accuracy across uncertainty deciles (before or after customization), and the distribution of predicted pruning ratios across uncertainty groups.

\textbf{Obs.4: List-wise ranking-uncertainty serves as an effective proxy for instance complexity.}
In Figure~\ref{fig:relation}(a), we visualize the relationship between ranking uncertainty and recommendation performance across ten uncertainty deciles on CD dataset with SASRec backbone. Observing the gray curve (before customization), we find a clear negative correlation: instances with lower uncertainty (D1 to D3) consistently achieve higher NDCG scores, while high-uncertainty instances (D8 to D10) exhibit clearly degraded performance. This empirically validates the theoretical argument in Section~\ref{sec:theory} that ranking uncertainty serves as a principled, data-driven proxy for instance complexity.
 
\textbf{Obs.5: RACE intelligently allocates resources based on uncertainty-aware threshold prediction and benefits instances of varying complexity through uncertainty-guided dual-space customization.}
Comparing performance before (gray line) and after (orange line) customization in Figure~\ref{fig:relation}(a) reveals that RACE improves recommendation accuracy for most uncertainty groups and reduces the performance gaps between them. High-uncertainty instances show the largest observed gains from additional capacity allocation, while low-uncertainty instances maintain comparable performance with reduced computation costs, demonstrating that RACE successfully addresses the Capacity-Complexity Mismatch through uncertainty-guided multi-level customization.

To further verify whether RACE learns to allocate resources intelligently, Figure~\ref{fig:relation}(b) shows the predicted channel pruning ratios across five uncertainty percentile groups (0-20\%, 20-40\%, 40-60\%, 60-80\%, 80-100\%), where boxes span the 25th to 75th percentiles, lines mark the medians, and whiskers extend to the 10th and 90th percentiles. The results reveal a clear downward trend: the negative correlation (Pearson's $r=-0.319$, Spearman's $\rho=-0.496$) and the fitted trend line confirm that RACE automatically learns to preserve more capacity for complex instances while aggressively compressing simpler ones, validating the effectiveness of our uncertainty-guided resource allocation.

\subsection{Training Robustness Analysis (RQ4)}
\label{sec:stability}

\textbf{Obs.6: RACE exhibits stable training dynamics and remains robust across hyperparameter choices.}

\begin{figure}[t]
    \vspace{4mm}
    \centering
    \includegraphics[width=0.92\linewidth]{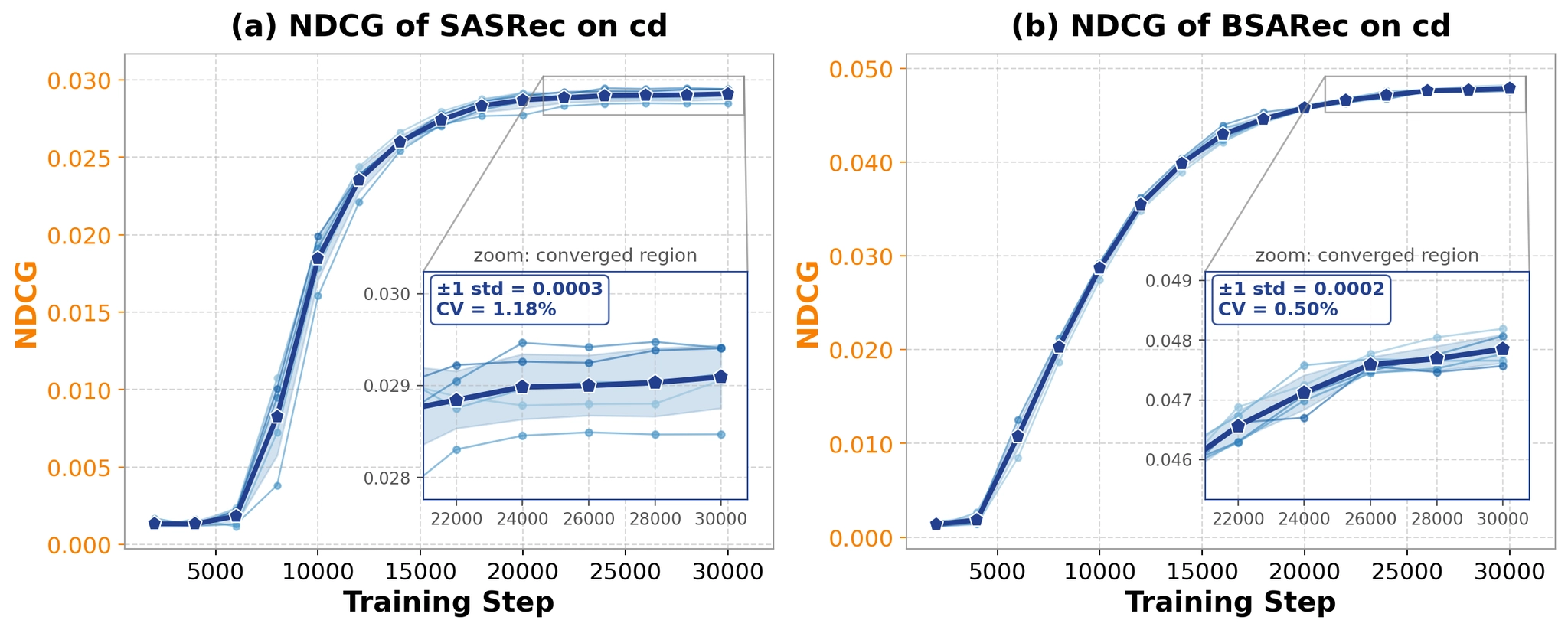}
    \vspace{-1mm}
    \caption{Training Curve of RACE over Five Seeds.}
    \label{fig:train_process}
\end{figure}

\looseness=-1
\textbf{\textit{Training stability analysis.} }Figure~\ref{fig:train_process} plots RACE's training curves on Amazon-CD over five random seeds with SASRec and BSARec. The tight $\pm$1 std bands and low coefficients of variation in the converged region (1.18\% and 0.50\%, respectively) confirm stable optimization across random initializations.

\begin{figure}[t]
    \centering
    \includegraphics[width=0.95\linewidth]{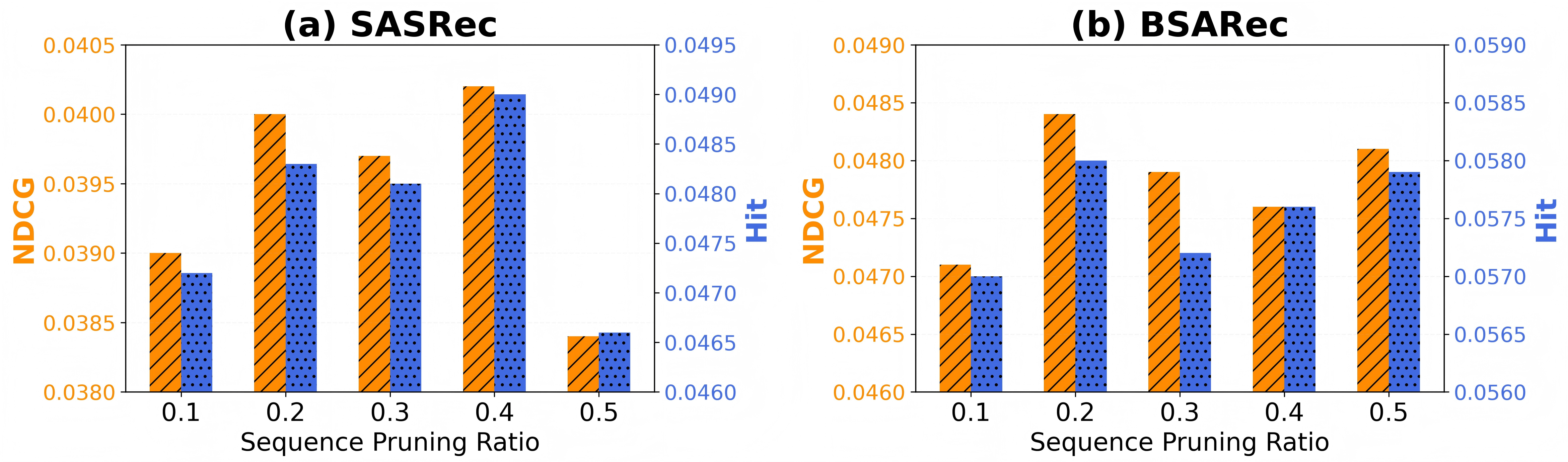}
    \caption{Sensitivity Analysis of Sequence Pruning Ratio.}
    \label{fig:sensitivity2}
\end{figure}

\begin{figure}[t]
    \centering
    \includegraphics[width=0.95\linewidth]{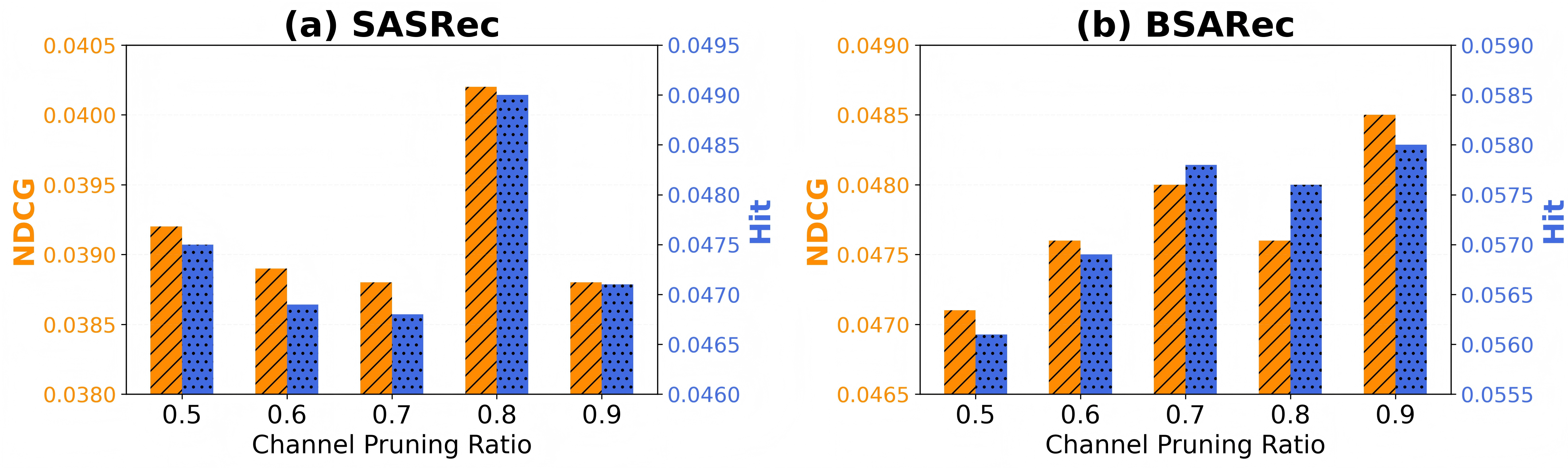}
    \vspace{-1mm}
    \caption{Sensitivity Analysis of Channel Pruning Ratio.}
    \label{fig:sensitivity1}
\end{figure}

\looseness=-1
\textbf{\textit{Pruning ratio sensitivity analysis.}} Figures~\ref{fig:sensitivity2} and~\ref{fig:sensitivity1} examine the layer-wise sequence and channel pruning ratios, respectively. Sequence pruning peaks at 0.4 on SASRec, while BSARec varies only mildly from 0.2 to 0.5. Channel pruning favors the high-ratio regime: 0.8 performs best on SASRec, and BSARec remains strong from 0.7 to 0.9. We therefore use 0.4 and 0.8 as shared defaults to maintain strong performance across both backbones while retaining substantial pruning.

\textbf{\textit{Loss-coefficient sensitivity analysis.}} On BSARec/Amazon-CD, the default ($\lambda_1=1$, $\lambda_2=0.5$) yields the best observed result; NDCG varies by at most 2.1\%/6.4\% across the evaluated $\lambda_1$/$\lambda_2$ ranges, respectively.

\vspace{-1mm}
\subsection{Inference Efficiency and Adaptivity Analysis (RQ5)}
\label{sec:deployment}

We evaluate inference efficiency under typical resource budgets and adaptivity to test-time constraints that differ from training.

\vspace{-0.5mm}
\subsubsection{Efficiency Analysis}
\looseness=-1
To evaluate efficiency under resource constraints, we simulate local reranking with a pool of 50 candidates, prefetched from the cloud, following established on-device settings~\cite{gong2020edgerec,gong2022real,fu2024DIET}. On SASRec, Table~\ref{tab:efficiency_analysis} reports backbone and auxiliary FLOPs together with latency profiled on a 4-core CPU.

\begin{table}[!h]
\caption{Efficiency analysis on the SASRec backbone.}
\vspace{-1mm}
\label{tab:efficiency_analysis}
\centering
{\footnotesize
\setlength{\tabcolsep}{3pt}
\renewcommand{\arraystretch}{1.0}
\begin{tabular*}{\linewidth}{@{\extracolsep{\fill}}lccc@{}}
\toprule
\textbf{Method} & \textbf{Backbone FLOPs} $\downarrow$ & \textbf{Other FLOPs} $\downarrow$ & \textbf{Latency} $\downarrow$ \\
\midrule
\texttt{Finetune} & 1.8432 & 3.6864 & 9.732 \\
\texttt{DIET} & 0.9010 & 3.0597 & 10.518 \\
\textbf{\texttt{RACE}} & \textbf{0.7549} & \textbf{0.0490} & \textbf{5.780} \\
\bottomrule
\end{tabular*}}
\end{table}

\textbf{Obs.7: RACE achieves efficient inference with negligible auxiliary cost through intrinsic feature reuse.}
\looseness=-1
Compared with Finetune, RACE reduces backbone FLOPs from 1.8432 to 0.7549 MFLOPs, a 59.0\% reduction. Its 0.0490-MFLOP auxiliary overhead is 62$\times$ lower than DIET's (3.0597 MFLOPs) and 75$\times$ lower than Finetune's backpropagation cost. At 5.780 ms per CPU inference, RACE reduces latency by 40.6\% and 45.0\% relative to Finetune and DIET, respectively.
 
\vspace{-0.5mm}
\subsubsection{Adaptivity Analysis}
\looseness=-1
Table~\ref{tab:adaptivity_analysis} evaluates RACE on Douban and CD with SASRec under test-time constraints different from training.

\begin{table}[!h]
\caption{Adaptivity analysis under varying test-time constraints on the SASRec backbone.}
\vspace{-1mm}
\label{tab:adaptivity_analysis}
\centering
\scriptsize
\renewcommand{\arraystretch}{0.86}
\resizebox{0.98\linewidth}{!}{
    \begin{tabular}{@{}llccc@{}}
    \toprule
     \textbf{Dataset} & \textbf{Test-time Constraint} & \textbf{NDCG} $\uparrow$ & \textbf{Hit} $\uparrow$ & \textbf{FLOPs} $\downarrow$ \\
     \midrule
     \multirow{3}{*}{Douban}
       & Stricter than training & 0.0259 & 0.0337 & 0.6266 \\
       & Same as training       & 0.0261 & 0.0337 & 0.7928 \\
       & Looser than training   & 0.0262 & 0.0341 & 1.0408 \\
     \midrule
     \multirow{3}{*}{CD}
       & Stricter than training & 0.0287 & 0.0328 & 0.6718 \\
       & Same as training       & 0.0288 & 0.0329 & 0.7549 \\
       & Looser than training   & 0.0288 & 0.0330 & 0.8567 \\
    \bottomrule
    \end{tabular}
}
\vspace{0.5mm}
\end{table}

\textbf{Obs.8: RACE gracefully trades accuracy and compute under varying inference constraints.}
\looseness=-1
Under stricter constraints than training, RACE incurs only $0.8\%$ / $0.3\%$ NDCG drops on Douban and CD while reducing FLOPs by an additional $21.0\%$ / $11.0\%$. Under looser constraints, the extra budget yields consistent NDCG and Hit improvements across both datasets. Together, these results demonstrate that RACE learns a flexible resource allocation policy that can adapt to varying inference constraints without retraining.

\vspace{1mm}
\section{Conclusion}
\looseness=-1
In this paper, we identify the Capacity-Complexity Mismatch in efficient sequential recommendation and propose RACE, which moves resource-constrained local inference from static-capacity compression to dynamic, uncertainty-aware adaptation. Guided by self-supervised list-wise ranking uncertainty as a proactive resource signal, RACE performs multi-granularity slimming across parameter and data spaces via hypernetwork-style channel customization and importance-aware progressive filtering, with low overhead from intrinsic feature reuse. Extensive experiments show that RACE balances recommendation accuracy and inference efficiency across diverse resource-constrained scenarios. Beyond on-device compression, this label-free self-assessment generalizes to broader sequential recommendation tasks such as performance estimation and model refresh scheduling. In future work, we plan to extend this paradigm to generative recommenders, where ranking uncertainty could guide adaptive decoding depth and selective beam expansion.

\section{Ethical Considerations}
RACE does not require additional data collection beyond the interaction histories used by the underlying recommender and does not alter its recommendation objective. Practical deployments should follow the same data-governance, access-control, and quality-evaluation practices as the underlying system.

\bibliographystyle{ACM-Reference-Format}
\bibliography{sample-base}


\begin{thebibliography}{62}


\ifx \showCODEN    \undefined \def \showCODEN     #1{\unskip}     \fi
\ifx \showISBNx    \undefined \def \showISBNx     #1{\unskip}     \fi
\ifx \showISBNxiii \undefined \def \showISBNxiii  #1{\unskip}     \fi
\ifx \showISSN     \undefined \def \showISSN      #1{\unskip}     \fi
\ifx \showLCCN     \undefined \def \showLCCN      #1{\unskip}     \fi
\ifx \shownote     \undefined \def \shownote      #1{#1}          \fi
\ifx \showarticletitle \undefined \def \showarticletitle #1{#1}   \fi
\ifx \showURL      \undefined \def \showURL       {\relax}        \fi
\providecommand\bibfield[2]{#2}
\providecommand\bibinfo[2]{#2}
\providecommand\natexlab[1]{#1}
\providecommand\showeprint[2][]{arXiv:#2}
\makeatletter
\@ifundefined{NAT@parse@date}{}{\let\NAT@parse@date@orig\NAT@parse@date}
\@ifundefined{NAT@parse@date}{}{\def\NAT@parse@date#1#2#3#4#5#6@@{\NAT@parse@date@orig#1#2#3#4#5#6@@\def\NAT@tempyear{0000}\def\NAT@tempexlab{{?}}\ifx\NAT@year\NAT@tempyear\ifx\NAT@exlab\NAT@tempexlab\def\NAT@date{[n.\,d.]}\else\edef\NAT@date{[n.\,d.]\NAT@exlab}\fi\fi}}
\makeatother

\bibitem[Bai et~al\mbox{.}(2022)]%
        {bai2022parameter}
\bibfield{author}{\bibinfo{person}{Yue Bai}, \bibinfo{person}{Huan Wang}, \bibinfo{person}{Xu Ma}, \bibinfo{person}{Yitian Zhang}, \bibinfo{person}{Zhiqiang Tao}, {and} \bibinfo{person}{Yun Fu}.} \bibinfo{year}{2022}\natexlab{}.
\newblock \showarticletitle{Parameter-efficient masking networks}.
\newblock \bibinfo{journal}{\emph{Advances in Neural Information Processing Systems}}  \bibinfo{volume}{35} (\bibinfo{year}{2022}), \bibinfo{pages}{10217--10229}.
\newblock


\bibitem[Bengio et~al\mbox{.}(2009)]%
        {bengio2009curriculum}
\bibfield{author}{\bibinfo{person}{Yoshua Bengio}, \bibinfo{person}{J{\'e}r{\^o}me Louradour}, \bibinfo{person}{Ronan Collobert}, {and} \bibinfo{person}{Jason Weston}.} \bibinfo{year}{2009}\natexlab{}.
\newblock \showarticletitle{Curriculum learning}. In \bibinfo{booktitle}{\emph{Proceedings of the 26th annual international conference on machine learning}}. \bibinfo{pages}{41--48}.
\newblock


\bibitem[Blundell et~al\mbox{.}(2015)]%
        {blundell2015weight}
\bibfield{author}{\bibinfo{person}{Charles Blundell}, \bibinfo{person}{Julien Cornebise}, \bibinfo{person}{Koray Kavukcuoglu}, {and} \bibinfo{person}{Daan Wierstra}.} \bibinfo{year}{2015}\natexlab{}.
\newblock \showarticletitle{Weight uncertainty in neural network}. In \bibinfo{booktitle}{\emph{International conference on machine learning}}. PMLR, \bibinfo{pages}{1613--1622}.
\newblock


\bibitem[Cai et~al\mbox{.}(2021)]%
        {cai2021category}
\bibfield{author}{\bibinfo{person}{Renqin Cai}, \bibinfo{person}{Jibang Wu}, \bibinfo{person}{Aidan San}, \bibinfo{person}{Chong Wang}, {and} \bibinfo{person}{Hongning Wang}.} \bibinfo{year}{2021}\natexlab{}.
\newblock \showarticletitle{Category-aware collaborative sequential recommendation}. In \bibinfo{booktitle}{\emph{Proceedings of the 44th international ACM SIGIR conference on research and development in information retrieval}}. \bibinfo{pages}{388--397}.
\newblock


\bibitem[Chang et~al\mbox{.}(2017)]%
        {chang2017streaming}
\bibfield{author}{\bibinfo{person}{Shiyu Chang}, \bibinfo{person}{Yang Zhang}, \bibinfo{person}{Jiliang Tang}, \bibinfo{person}{Dawei Yin}, \bibinfo{person}{Yi Chang}, \bibinfo{person}{Mark~A Hasegawa-Johnson}, {and} \bibinfo{person}{Thomas~S Huang}.} \bibinfo{year}{2017}\natexlab{}.
\newblock \showarticletitle{Streaming recommender systems}. In \bibinfo{booktitle}{\emph{Proceedings of the 26th international conference on world wide web}}. \bibinfo{pages}{381--389}.
\newblock


\bibitem[Chen et~al\mbox{.}(2019)]%
        {chen2019you}
\bibfield{author}{\bibinfo{person}{Zhourong Chen}, \bibinfo{person}{Yang Li}, \bibinfo{person}{Samy Bengio}, {and} \bibinfo{person}{Si Si}.} \bibinfo{year}{2019}\natexlab{}.
\newblock \showarticletitle{You look twice: Gaternet for dynamic filter selection in cnns}. In \bibinfo{booktitle}{\emph{Proceedings of the IEEE/CVF Conference on Computer Vision and Pattern Recognition}}. \bibinfo{pages}{9172--9180}.
\newblock


\bibitem[Coscrato and Bridge(2023)]%
        {10.1145/3584021}
\bibfield{author}{\bibinfo{person}{Victor Coscrato} {and} \bibinfo{person}{Derek Bridge}.} \bibinfo{year}{2023}\natexlab{}.
\newblock \showarticletitle{Estimating and Evaluating the Uncertainty of Rating Predictions and Top-n Recommendations in Recommender Systems}.
\newblock \bibinfo{journal}{\emph{ACM Trans. Recomm. Syst.}} \bibinfo{volume}{1}, \bibinfo{number}{2}, Article \bibinfo{articleno}{7} (\bibinfo{date}{April} \bibinfo{year}{2023}), \bibinfo{numpages}{34}~pages.
\newblock
\href{https://doi.org/10.1145/3584021}{doi:\nolinkurl{10.1145/3584021}}


\bibitem[Dai et~al\mbox{.}(2024)]%
        {dai2024modeling}
\bibfield{author}{\bibinfo{person}{Sunhao Dai}, \bibinfo{person}{Ninglu Shao}, \bibinfo{person}{Jieming Zhu}, \bibinfo{person}{Xiao Zhang}, \bibinfo{person}{Zhenhua Dong}, \bibinfo{person}{Jun Xu}, \bibinfo{person}{Quanyu Dai}, {and} \bibinfo{person}{Ji-Rong Wen}.} \bibinfo{year}{2024}\natexlab{}.
\newblock \showarticletitle{Modeling user attention in music recommendation}. In \bibinfo{booktitle}{\emph{2024 IEEE 40th International Conference on Data Engineering (ICDE)}}. IEEE, \bibinfo{pages}{761--774}.
\newblock


\bibitem[Du et~al\mbox{.}(2023)]%
        {du2023idnp}
\bibfield{author}{\bibinfo{person}{Jing Du}, \bibinfo{person}{Zesheng Ye}, \bibinfo{person}{Bin Guo}, \bibinfo{person}{Zhiwen Yu}, {and} \bibinfo{person}{Lina Yao}.} \bibinfo{year}{2023}\natexlab{}.
\newblock \showarticletitle{Idnp: Interest dynamics modeling using generative neural processes for sequential recommendation}. In \bibinfo{booktitle}{\emph{Proceedings of the Sixteenth ACM International Conference on web search and data mining}}. \bibinfo{pages}{481--489}.
\newblock


\bibitem[Fu et~al\mbox{.}(2026a)]%
        {rankgr2026}
\bibfield{author}{\bibinfo{person}{Kairui Fu}, \bibinfo{person}{Changfa Wu}, \bibinfo{person}{Kun Yuan}, \bibinfo{person}{Binbin Cao}, \bibinfo{person}{Dunxian Huang}, \bibinfo{person}{Yuliang Yan}, \bibinfo{person}{Junjun Zheng}, \bibinfo{person}{Jianning Zhang}, \bibinfo{person}{Silu Zhou}, \bibinfo{person}{Jian Wu}, {and} \bibinfo{person}{Kun Kuang}.} \bibinfo{year}{2026}\natexlab{a}.
\newblock \showarticletitle{{RankGR}: Rank-Enhanced Generative Retrieval with Listwise Direct Preference Optimization in Recommendation}.
\newblock \bibinfo{journal}{\emph{arXiv preprint arXiv:2602.08575}} (\bibinfo{year}{2026}).
\newblock


\bibitem[Fu et~al\mbox{.}(2024)]%
        {fu2024DIET}
\bibfield{author}{\bibinfo{person}{Kairui Fu}, \bibinfo{person}{Shengyu Zhang}, \bibinfo{person}{Zheqi Lv}, \bibinfo{person}{Jingyuan Chen}, {and} \bibinfo{person}{Jiwei Li}.} \bibinfo{year}{2024}\natexlab{}.
\newblock \showarticletitle{DIET: Customized slimming for incompatible networks in sequential recommendation}. In \bibinfo{booktitle}{\emph{Proceedings of the 30th ACM SIGKDD Conference on Knowledge Discovery and Data Mining}}. \bibinfo{pages}{816--826}.
\newblock


\bibitem[Fu et~al\mbox{.}(2026b)]%
        {forge2026}
\bibfield{author}{\bibinfo{person}{Kairui Fu}, \bibinfo{person}{Tao Zhang}, \bibinfo{person}{Shuwen Xiao}, \bibinfo{person}{Ziyang Wang}, \bibinfo{person}{Xinming Zhang}, \bibinfo{person}{Chenchi Zhang}, \bibinfo{person}{Yuliang Yan}, \bibinfo{person}{Junjun Zheng}, \bibinfo{person}{Xiangheng Kong}, \bibinfo{person}{Shengyu Zhang}, \bibinfo{person}{Kun Kuang}, {and} \bibinfo{person}{Yuning Jiang}.} \bibinfo{year}{2026}\natexlab{b}.
\newblock \showarticletitle{{FORGE}: Forming Semantic Identifiers for Generative Retrieval in Industrial Datasets}. In \bibinfo{booktitle}{\emph{Proceedings of the 32nd ACM SIGKDD Conference on Knowledge Discovery and Data Mining V.2}}. \bibinfo{pages}{8858--8869}.
\newblock


\bibitem[Gal and Ghahramani(2016)]%
        {gal2016dropout}
\bibfield{author}{\bibinfo{person}{Yarin Gal} {and} \bibinfo{person}{Zoubin Ghahramani}.} \bibinfo{year}{2016}\natexlab{}.
\newblock \showarticletitle{Dropout as a bayesian approximation: Representing model uncertainty in deep learning}. In \bibinfo{booktitle}{\emph{international conference on machine learning}}. PMLR, \bibinfo{pages}{1050--1059}.
\newblock


\bibitem[Gong et~al\mbox{.}(2022)]%
        {gong2022real}
\bibfield{author}{\bibinfo{person}{Xudong Gong}, \bibinfo{person}{Qinlin Feng}, \bibinfo{person}{Yuan Zhang}, \bibinfo{person}{Jiangling Qin}, \bibinfo{person}{Weijie Ding}, \bibinfo{person}{Biao Li}, \bibinfo{person}{Peng Jiang}, {and} \bibinfo{person}{Kun Gai}.} \bibinfo{year}{2022}\natexlab{}.
\newblock \showarticletitle{Real-time short video recommendation on mobile devices}. In \bibinfo{booktitle}{\emph{Proceedings of the 31st ACM international conference on information \& knowledge management}}. \bibinfo{pages}{3103--3112}.
\newblock


\bibitem[Gong et~al\mbox{.}(2020)]%
        {gong2020edgerec}
\bibfield{author}{\bibinfo{person}{Yu Gong}, \bibinfo{person}{Ziwen Jiang}, \bibinfo{person}{Yufei Feng}, \bibinfo{person}{Binbin Hu}, \bibinfo{person}{Kaiqi Zhao}, \bibinfo{person}{Qingwen Liu}, {and} \bibinfo{person}{Wenwu Ou}.} \bibinfo{year}{2020}\natexlab{}.
\newblock \showarticletitle{EdgeRec: recommender system on edge in Mobile Taobao}. In \bibinfo{booktitle}{\emph{Proceedings of the 29th ACM International Conference on Information \& Knowledge Management}}. \bibinfo{pages}{2477--2484}.
\newblock


\bibitem[Goyal et~al\mbox{.}(2020)]%
        {goyal2020power}
\bibfield{author}{\bibinfo{person}{Saurabh Goyal}, \bibinfo{person}{Anamitra~Roy Choudhury}, \bibinfo{person}{Saurabh Raje}, \bibinfo{person}{Venkatesan Chakaravarthy}, \bibinfo{person}{Yogish Sabharwal}, {and} \bibinfo{person}{Ashish Verma}.} \bibinfo{year}{2020}\natexlab{}.
\newblock \showarticletitle{Power-bert: Accelerating bert inference via progressive word-vector elimination}. In \bibinfo{booktitle}{\emph{International Conference on Machine Learning}}. PMLR, \bibinfo{pages}{3690--3699}.
\newblock


\bibitem[Guan et~al\mbox{.}(2022)]%
        {guan2022transkimmer}
\bibfield{author}{\bibinfo{person}{Yue Guan}, \bibinfo{person}{Zhengyi Li}, \bibinfo{person}{Jingwen Leng}, \bibinfo{person}{Zhouhan Lin}, {and} \bibinfo{person}{Minyi Guo}.} \bibinfo{year}{2022}\natexlab{}.
\newblock \showarticletitle{Transkimmer: Transformer learns to layer-wise skim}.
\newblock \bibinfo{journal}{\emph{arXiv preprint arXiv:2205.07324}} (\bibinfo{year}{2022}).
\newblock


\bibitem[Guo et~al\mbox{.}(2017)]%
        {guo2017calibration}
\bibfield{author}{\bibinfo{person}{Chuan Guo}, \bibinfo{person}{Geoff Pleiss}, \bibinfo{person}{Yu Sun}, {and} \bibinfo{person}{Kilian~Q Weinberger}.} \bibinfo{year}{2017}\natexlab{}.
\newblock \showarticletitle{On calibration of modern neural networks}. In \bibinfo{booktitle}{\emph{International conference on machine learning}}. PMLR, \bibinfo{pages}{1321--1330}.
\newblock


\bibitem[Harrison et~al\mbox{.}(2024)]%
        {harrison2024variational}
\bibfield{author}{\bibinfo{person}{James Harrison}, \bibinfo{person}{John Willes}, {and} \bibinfo{person}{Jasper Snoek}.} \bibinfo{year}{2024}\natexlab{}.
\newblock \showarticletitle{Variational Bayesian last layers}.
\newblock \bibinfo{journal}{\emph{arXiv preprint arXiv:2404.11599}} (\bibinfo{year}{2024}).
\newblock


\bibitem[Heuss et~al\mbox{.}(2023)]%
        {heusspredictive2023}
\bibfield{author}{\bibinfo{person}{Maria Heuss}, \bibinfo{person}{Daniel Cohen}, \bibinfo{person}{Masoud Mansoury}, \bibinfo{person}{Maarten~de Rijke}, {and} \bibinfo{person}{Carsten Eickhoff}.} \bibinfo{year}{2023}\natexlab{}.
\newblock \showarticletitle{Predictive uncertainty-based bias mitigation in ranking}. In \bibinfo{booktitle}{\emph{Proceedings of the 32nd ACM International Conference on Information and Knowledge Management}}. \bibinfo{pages}{762--772}.
\newblock


\bibitem[H{\"u}llermeier and Waegeman(2021)]%
        {hullermeier2021aleatoric}
\bibfield{author}{\bibinfo{person}{Eyke H{\"u}llermeier} {and} \bibinfo{person}{Willem Waegeman}.} \bibinfo{year}{2021}\natexlab{}.
\newblock \showarticletitle{Aleatoric and epistemic uncertainty in machine learning: An introduction to concepts and methods}.
\newblock \bibinfo{journal}{\emph{Machine learning}} \bibinfo{volume}{110}, \bibinfo{number}{3} (\bibinfo{year}{2021}), \bibinfo{pages}{457--506}.
\newblock


\bibitem[Jeon et~al\mbox{.}(2024)]%
        {jeon2024calibration}
\bibfield{author}{\bibinfo{person}{Hyunsik Jeon}, \bibinfo{person}{Se-eun Yoon}, {and} \bibinfo{person}{Julian McAuley}.} \bibinfo{year}{2024}\natexlab{}.
\newblock \showarticletitle{Calibration-disentangled learning and relevance-prioritized reranking for calibrated sequential recommendation}. In \bibinfo{booktitle}{\emph{Proceedings of the 33rd ACM International Conference on Information and Knowledge Management}}. \bibinfo{pages}{973--982}.
\newblock


\bibitem[Kang and McAuley(2018)]%
        {kang2018self}
\bibfield{author}{\bibinfo{person}{Wang-Cheng Kang} {and} \bibinfo{person}{Julian McAuley}.} \bibinfo{year}{2018}\natexlab{}.
\newblock \showarticletitle{Self-attentive sequential recommendation}. In \bibinfo{booktitle}{\emph{2018 IEEE international conference on data mining (ICDM)}}. IEEE, \bibinfo{pages}{197--206}.
\newblock


\bibitem[Kendall and Gal(2017)]%
        {kendall2017uncertainties}
\bibfield{author}{\bibinfo{person}{Alex Kendall} {and} \bibinfo{person}{Yarin Gal}.} \bibinfo{year}{2017}\natexlab{}.
\newblock \showarticletitle{What uncertainties do we need in bayesian deep learning for computer vision?}
\newblock \bibinfo{journal}{\emph{Advances in neural information processing systems}}  \bibinfo{volume}{30} (\bibinfo{year}{2017}).
\newblock


\bibitem[Khan et~al\mbox{.}(2019)]%
        {khan2019edge}
\bibfield{author}{\bibinfo{person}{Wazir~Zada Khan}, \bibinfo{person}{Ejaz Ahmed}, \bibinfo{person}{Saqib Hakak}, \bibinfo{person}{Ibrar Yaqoob}, {and} \bibinfo{person}{Arif Ahmed}.} \bibinfo{year}{2019}\natexlab{}.
\newblock \showarticletitle{Edge computing: A survey}.
\newblock \bibinfo{journal}{\emph{Future Generation Computer Systems}}  \bibinfo{volume}{97} (\bibinfo{year}{2019}), \bibinfo{pages}{219--235}.
\newblock


\bibitem[Kim et~al\mbox{.}(2022)]%
        {kim2022learned}
\bibfield{author}{\bibinfo{person}{Sehoon Kim}, \bibinfo{person}{Sheng Shen}, \bibinfo{person}{David Thorsley}, \bibinfo{person}{Amir Gholami}, \bibinfo{person}{Woosuk Kwon}, \bibinfo{person}{Joseph Hassoun}, {and} \bibinfo{person}{Kurt Keutzer}.} \bibinfo{year}{2022}\natexlab{}.
\newblock \showarticletitle{Learned token pruning for transformers}. In \bibinfo{booktitle}{\emph{Proceedings of the 28th ACM SIGKDD conference on knowledge discovery and data mining}}. \bibinfo{pages}{784--794}.
\newblock


\bibitem[Lakshminarayanan et~al\mbox{.}(2017)]%
        {lakshminarayanan2017simple}
\bibfield{author}{\bibinfo{person}{Balaji Lakshminarayanan}, \bibinfo{person}{Alexander Pritzel}, {and} \bibinfo{person}{Charles Blundell}.} \bibinfo{year}{2017}\natexlab{}.
\newblock \showarticletitle{Simple and scalable predictive uncertainty estimation using deep ensembles}.
\newblock \bibinfo{journal}{\emph{Advances in neural information processing systems}}  \bibinfo{volume}{30} (\bibinfo{year}{2017}).
\newblock


\bibitem[Lee et~al\mbox{.}(2018)]%
        {lee2018snip}
\bibfield{author}{\bibinfo{person}{Namhoon Lee}, \bibinfo{person}{Thalaiyasingam Ajanthan}, {and} \bibinfo{person}{Philip~HS Torr}.} \bibinfo{year}{2018}\natexlab{}.
\newblock \showarticletitle{Snip: Single-shot network pruning based on connection sensitivity}.
\newblock \bibinfo{journal}{\emph{arXiv preprint arXiv:1810.02340}} (\bibinfo{year}{2018}).
\newblock


\bibitem[Lei et~al\mbox{.}(2021)]%
        {lei2021semi}
\bibfield{author}{\bibinfo{person}{Chenyi Lei}, \bibinfo{person}{Yong Liu}, \bibinfo{person}{Lingzi Zhang}, \bibinfo{person}{Guoxin Wang}, \bibinfo{person}{Haihong Tang}, \bibinfo{person}{Houqiang Li}, {and} \bibinfo{person}{Chunyan Miao}.} \bibinfo{year}{2021}\natexlab{}.
\newblock \showarticletitle{Semi: A sequential multi-modal information transfer network for e-commerce micro-video recommendations}. In \bibinfo{booktitle}{\emph{Proceedings of the 27th ACM SIGKDD Conference on Knowledge Discovery \& Data Mining}}. \bibinfo{pages}{3161--3171}.
\newblock


\bibitem[Li et~al\mbox{.}(2025)]%
        {li2025recommendersselfawarelabelfreerecommendation}
\bibfield{author}{\bibinfo{person}{Jiayu Li}, \bibinfo{person}{Ziyi Ye}, \bibinfo{person}{Guohao Jian}, \bibinfo{person}{Zhiqiang Guo}, \bibinfo{person}{Weizhi Ma}, \bibinfo{person}{Qingyao Ai}, {and} \bibinfo{person}{Min Zhang}.} \bibinfo{year}{2025}\natexlab{}.
\newblock \showarticletitle{Are Recommenders Self-Aware? Label-Free Recommendation Performance Estimation via Model Uncertainty}.
\newblock \bibinfo{journal}{\emph{arXiv preprint arXiv:2507.23208}} (\bibinfo{year}{2025}).
\newblock


\bibitem[Liu et~al\mbox{.}(2026)]%
        {liu2026accelerating}
\bibfield{author}{\bibinfo{person}{Qijiong Liu}, \bibinfo{person}{Lu Fan}, \bibinfo{person}{Zhongzhou Liu}, \bibinfo{person}{Xiaoyu Dong}, \bibinfo{person}{Yuankai Luo}, \bibinfo{person}{Guoyuan An}, \bibinfo{person}{Nuo Chen}, \bibinfo{person}{Wei Guo}, \bibinfo{person}{Yong Liu}, {and} \bibinfo{person}{Xiao-Ming Wu}.} \bibinfo{year}{2026}\natexlab{}.
\newblock \showarticletitle{Accelerating Generative Recommendation via Simple Categorical User Sequence Compression}. In \bibinfo{booktitle}{\emph{Proceedings of the Nineteenth ACM International Conference on Web Search and Data Mining}}. \bibinfo{pages}{1195--1199}.
\newblock


\bibitem[Liu et~al\mbox{.}(2025)]%
        {liu2025chord}
\bibfield{author}{\bibinfo{person}{Tianqi Liu}, \bibinfo{person}{Kairui Fu}, \bibinfo{person}{Shengyu Zhang}, \bibinfo{person}{Wenyan Fan}, \bibinfo{person}{Zhaocheng Du}, \bibinfo{person}{Jieming Zhu}, \bibinfo{person}{Fan Wu}, {and} \bibinfo{person}{Fei Wu}.} \bibinfo{year}{2025}\natexlab{}.
\newblock \showarticletitle{{CHORD}: Customizing Hybrid-precision On-device Model for Sequential Recommendation with Device-cloud Collaboration}. In \bibinfo{booktitle}{\emph{Proceedings of the 33rd ACM International Conference on Multimedia}}. \bibinfo{publisher}{Association for Computing Machinery}, \bibinfo{address}{New York, NY, USA}, \bibinfo{pages}{6252--6261}.
\newblock
\href{https://doi.org/10.1145/3746027.3755632}{doi:\nolinkurl{10.1145/3746027.3755632}}


\bibitem[Lv et~al\mbox{.}(2025)]%
        {lsc4rec2025}
\bibfield{author}{\bibinfo{person}{Zheqi Lv}, \bibinfo{person}{Tianyu Zhan}, \bibinfo{person}{Wenjie Wang}, \bibinfo{person}{Xinyu Lin}, \bibinfo{person}{Shengyu Zhang}, \bibinfo{person}{Wenqiao Zhang}, \bibinfo{person}{Jiwei Li}, \bibinfo{person}{Kun Kuang}, {and} \bibinfo{person}{Fei Wu}.} \bibinfo{year}{2025}\natexlab{}.
\newblock \showarticletitle{Collaboration of Large Language Models and Small Recommendation Models for Device-Cloud Recommendation}. In \bibinfo{booktitle}{\emph{Proceedings of the 31st ACM SIGKDD Conference on Knowledge Discovery and Data Mining V.1}}. \bibinfo{pages}{962--973}.
\newblock


\bibitem[Lv et~al\mbox{.}(2024)]%
        {lv2024intelligent}
\bibfield{author}{\bibinfo{person}{Zheqi Lv}, \bibinfo{person}{Wenqiao Zhang}, \bibinfo{person}{Zhengyu Chen}, \bibinfo{person}{Shengyu Zhang}, {and} \bibinfo{person}{Kun Kuang}.} \bibinfo{year}{2024}\natexlab{}.
\newblock \showarticletitle{Intelligent model update strategy for sequential recommendation}. In \bibinfo{booktitle}{\emph{Proceedings of the ACM Web Conference 2024}}. \bibinfo{pages}{3117--3128}.
\newblock


\bibitem[Mairittha et~al\mbox{.}(2020)]%
        {mairittha2020improving}
\bibfield{author}{\bibinfo{person}{Nattaya Mairittha}, \bibinfo{person}{Tittaya Mairittha}, {and} \bibinfo{person}{Sozo Inoue}.} \bibinfo{year}{2020}\natexlab{}.
\newblock \showarticletitle{Improving activity data collection with on-device personalization using fine-tuning}. In \bibinfo{booktitle}{\emph{UbiComp/ISWC}}. \bibinfo{pages}{255--260}.
\newblock


\bibitem[Qian et~al\mbox{.}(2022)]%
        {qian2022intelligent}
\bibfield{author}{\bibinfo{person}{Xufeng Qian}, \bibinfo{person}{Yue Xu}, \bibinfo{person}{Fuyu Lv}, \bibinfo{person}{Shengyu Zhang}, \bibinfo{person}{Ziwen Jiang}, \bibinfo{person}{Qingwen Liu}, \bibinfo{person}{Xiaoyi Zeng}, \bibinfo{person}{Tat-Seng Chua}, {and} \bibinfo{person}{Fei Wu}.} \bibinfo{year}{2022}\natexlab{}.
\newblock \showarticletitle{Intelligent request strategy design in recommender system}. In \bibinfo{booktitle}{\emph{Proceedings of the 28th ACM SIGKDD Conference on Knowledge Discovery and Data Mining}}. \bibinfo{pages}{3772--3782}.
\newblock


\bibitem[Shao et~al\mbox{.}(2025)]%
        {shao2025dtrec}
\bibfield{author}{\bibinfo{person}{Yifan Shao}, \bibinfo{person}{Peilin Zhou}, \bibinfo{person}{Shoujin Wang}, \bibinfo{person}{Weizhi Zhang}, \bibinfo{person}{Xu Cai}, {and} \bibinfo{person}{Sunghun Kim}.} \bibinfo{year}{2025}\natexlab{}.
\newblock \showarticletitle{DTRec: Learning Dynamic Reasoning Trajectories for Sequential Recommendation}.
\newblock \bibinfo{journal}{\emph{arXiv preprint arXiv:2512.14036}} (\bibinfo{year}{2025}).
\newblock


\bibitem[Shin et~al\mbox{.}(2024)]%
        {shin2024attentive}
\bibfield{author}{\bibinfo{person}{Yehjin Shin}, \bibinfo{person}{Jeongwhan Choi}, \bibinfo{person}{Hyowon Wi}, {and} \bibinfo{person}{Noseong Park}.} \bibinfo{year}{2024}\natexlab{}.
\newblock \showarticletitle{An attentive inductive bias for sequential recommendation beyond the self-attention}. In \bibinfo{booktitle}{\emph{Proceedings of the AAAI conference on artificial intelligence}}, Vol.~\bibinfo{volume}{38}. \bibinfo{pages}{8984--8992}.
\newblock


\bibitem[Sreenivasan et~al\mbox{.}(2022)]%
        {sreenivasan2022rare}
\bibfield{author}{\bibinfo{person}{Kartik Sreenivasan}, \bibinfo{person}{Jy-yong Sohn}, \bibinfo{person}{Liu Yang}, \bibinfo{person}{Matthew Grinde}, \bibinfo{person}{Alliot Nagle}, \bibinfo{person}{Hongyi Wang}, \bibinfo{person}{Eric Xing}, \bibinfo{person}{Kangwook Lee}, {and} \bibinfo{person}{Dimitris Papailiopoulos}.} \bibinfo{year}{2022}\natexlab{}.
\newblock \showarticletitle{Rare gems: Finding lottery tickets at initialization}.
\newblock \bibinfo{journal}{\emph{Advances in Neural Information Processing Systems}}  \bibinfo{volume}{35} (\bibinfo{year}{2022}), \bibinfo{pages}{14529--14540}.
\newblock


\bibitem[Sun et~al\mbox{.}(2019)]%
        {10.1145/3357384.3357895}
\bibfield{author}{\bibinfo{person}{Fei Sun}, \bibinfo{person}{Jun Liu}, \bibinfo{person}{Jian Wu}, \bibinfo{person}{Changhua Pei}, \bibinfo{person}{Xiao Lin}, \bibinfo{person}{Wenwu Ou}, {and} \bibinfo{person}{Peng Jiang}.} \bibinfo{year}{2019}\natexlab{}.
\newblock \showarticletitle{BERT4Rec: Sequential Recommendation with Bidirectional Encoder Representations from Transformer}. In \bibinfo{booktitle}{\emph{Proceedings of the 28th ACM International Conference on Information and Knowledge Management}} (Beijing, China) \emph{(\bibinfo{series}{CIKM '19})}. \bibinfo{publisher}{Association for Computing Machinery}, \bibinfo{address}{New York, NY, USA}, \bibinfo{pages}{1441–1450}.
\newblock
\showISBNx{9781450369763}
\href{https://doi.org/10.1145/3357384.3357895}{doi:\nolinkurl{10.1145/3357384.3357895}}


\bibitem[Tan et~al\mbox{.}(2021)]%
        {tan2021sparse}
\bibfield{author}{\bibinfo{person}{Qiaoyu Tan}, \bibinfo{person}{Jianwei Zhang}, \bibinfo{person}{Jiangchao Yao}, \bibinfo{person}{Ninghao Liu}, \bibinfo{person}{Jingren Zhou}, \bibinfo{person}{Hongxia Yang}, {and} \bibinfo{person}{Xia Hu}.} \bibinfo{year}{2021}\natexlab{}.
\newblock \showarticletitle{Sparse-interest network for sequential recommendation}. In \bibinfo{booktitle}{\emph{Proceedings of the 14th ACM international conference on web search and data mining}}. \bibinfo{pages}{598--606}.
\newblock


\bibitem[Thurstone(1927)]%
        {thurstone}
\bibfield{author}{\bibinfo{person}{Louis~Leon Thurstone}.} \bibinfo{year}{1927}\natexlab{}.
\newblock \showarticletitle{A Law of Comparative Judgment}.
\newblock \bibinfo{journal}{\emph{Psychological Review}} \bibinfo{volume}{34}, \bibinfo{number}{4} (\bibinfo{year}{1927}), \bibinfo{pages}{273--286}.
\newblock


\bibitem[Wang et~al\mbox{.}(2020b)]%
        {wang2020time}
\bibfield{author}{\bibinfo{person}{Jianling Wang}, \bibinfo{person}{Raphael Louca}, \bibinfo{person}{Diane Hu}, \bibinfo{person}{Caitlin Cellier}, \bibinfo{person}{James Caverlee}, {and} \bibinfo{person}{Liangjie Hong}.} \bibinfo{year}{2020}\natexlab{b}.
\newblock \showarticletitle{Time to shop for valentine's day: Shopping occasions and sequential recommendation in e-commerce}. In \bibinfo{booktitle}{\emph{Proceedings of the 13th international conference on web search and data mining}}. \bibinfo{pages}{645--653}.
\newblock


\bibitem[Wang et~al\mbox{.}(2021)]%
        {10.1007/s00778-021-00700-6}
\bibfield{author}{\bibinfo{person}{Qinyong Wang}, \bibinfo{person}{Hongzhi Yin}, \bibinfo{person}{Tong Chen}, \bibinfo{person}{Junliang Yu}, \bibinfo{person}{Alexander Zhou}, {and} \bibinfo{person}{Xiangliang Zhang}.} \bibinfo{year}{2021}\natexlab{}.
\newblock \showarticletitle{Fast-Adapting and Privacy-Preserving Federated Recommender System}.
\newblock \bibinfo{journal}{\emph{The VLDB Journal}} \bibinfo{volume}{31}, \bibinfo{number}{5} (\bibinfo{year}{2021}), \bibinfo{pages}{877–896}.
\newblock


\bibitem[Wang et~al\mbox{.}(2020a)]%
        {wang2020convergence}
\bibfield{author}{\bibinfo{person}{Xiaofei Wang}, \bibinfo{person}{Yiwen Han}, \bibinfo{person}{Victor~CM Leung}, \bibinfo{person}{Dusit Niyato}, \bibinfo{person}{Xueqiang Yan}, {and} \bibinfo{person}{Xu Chen}.} \bibinfo{year}{2020}\natexlab{a}.
\newblock \showarticletitle{Convergence of edge computing and deep learning: A comprehensive survey}.
\newblock \bibinfo{journal}{\emph{IEEE communications surveys \& tutorials}} \bibinfo{volume}{22}, \bibinfo{number}{2} (\bibinfo{year}{2020}), \bibinfo{pages}{869--904}.
\newblock


\bibitem[Xi et~al\mbox{.}(2023)]%
        {xi2023device}
\bibfield{author}{\bibinfo{person}{Yunjia Xi}, \bibinfo{person}{Weiwen Liu}, \bibinfo{person}{Yang Wang}, \bibinfo{person}{Ruiming Tang}, \bibinfo{person}{Weinan Zhang}, \bibinfo{person}{Yue Zhu}, \bibinfo{person}{Rui Zhang}, {and} \bibinfo{person}{Yong Yu}.} \bibinfo{year}{2023}\natexlab{}.
\newblock \showarticletitle{On-device integrated re-ranking with heterogeneous behavior modeling}. In \bibinfo{booktitle}{\emph{Proceedings of the 29th ACM SIGKDD Conference on Knowledge Discovery and Data Mining}}. \bibinfo{pages}{5225--5236}.
\newblock


\bibitem[Xia et~al\mbox{.}(2022)]%
        {xia2022device}
\bibfield{author}{\bibinfo{person}{Xin Xia}, \bibinfo{person}{Hongzhi Yin}, \bibinfo{person}{Junliang Yu}, \bibinfo{person}{Qinyong Wang}, \bibinfo{person}{Guandong Xu}, {and} \bibinfo{person}{Quoc Viet~Hung Nguyen}.} \bibinfo{year}{2022}\natexlab{}.
\newblock \showarticletitle{On-device next-item recommendation with self-supervised knowledge distillation}. In \bibinfo{booktitle}{\emph{Proceedings of the 45th International ACM SIGIR Conference on Research and Development in Information Retrieval}}. \bibinfo{pages}{546--555}.
\newblock


\bibitem[Yan et~al\mbox{.}(2022b)]%
        {yan2022apg}
\bibfield{author}{\bibinfo{person}{Bencheng Yan}, \bibinfo{person}{Pengjie Wang}, \bibinfo{person}{Kai Zhang}, \bibinfo{person}{Feng Li}, \bibinfo{person}{Hongbo Deng}, \bibinfo{person}{Jian Xu}, {and} \bibinfo{person}{Bo Zheng}.} \bibinfo{year}{2022}\natexlab{b}.
\newblock \showarticletitle{Apg: Adaptive parameter generation network for click-through rate prediction}.
\newblock \bibinfo{journal}{\emph{Advances in Neural Information Processing Systems}}  \bibinfo{volume}{35} (\bibinfo{year}{2022}), \bibinfo{pages}{24740--24752}.
\newblock


\bibitem[Yan et~al\mbox{.}(2022a)]%
        {yan2022device}
\bibfield{author}{\bibinfo{person}{Yikai Yan}, \bibinfo{person}{Chaoyue Niu}, \bibinfo{person}{Renjie Gu}, \bibinfo{person}{Fan Wu}, \bibinfo{person}{Shaojie Tang}, \bibinfo{person}{Lifeng Hua}, \bibinfo{person}{Chengfei Lyu}, {and} \bibinfo{person}{Guihai Chen}.} \bibinfo{year}{2022}\natexlab{a}.
\newblock \showarticletitle{On-device learning for model personalization with large-scale cloud-coordinated domain adaption}. In \bibinfo{booktitle}{\emph{Proceedings of the 28th ACM SIGKDD Conference on Knowledge Discovery and Data Mining}}. \bibinfo{pages}{2180--2190}.
\newblock


\bibitem[Yang et~al\mbox{.}(2024)]%
        {yangmitigating2023}
\bibfield{author}{\bibinfo{person}{Tao Yang}, \bibinfo{person}{Cuize Han}, \bibinfo{person}{Chen Luo}, \bibinfo{person}{Parth Gupta}, \bibinfo{person}{Jeff~M Phillips}, {and} \bibinfo{person}{Qingyao Ai}.} \bibinfo{year}{2024}\natexlab{}.
\newblock \showarticletitle{Mitigating exploitation bias in learning to rank with an uncertainty-aware empirical Bayes approach}. In \bibinfo{booktitle}{\emph{Proceedings of the ACM Web Conference 2024}}. \bibinfo{pages}{1486--1496}.
\newblock


\bibitem[Yao et~al\mbox{.}(2021)]%
        {yao2021device}
\bibfield{author}{\bibinfo{person}{Jiangchao Yao}, \bibinfo{person}{Feng Wang}, \bibinfo{person}{Kunyang Jia}, \bibinfo{person}{Bo Han}, \bibinfo{person}{Jingren Zhou}, {and} \bibinfo{person}{Hongxia Yang}.} \bibinfo{year}{2021}\natexlab{}.
\newblock \showarticletitle{Device-cloud collaborative learning for recommendation}. In \bibinfo{booktitle}{\emph{Proceedings of the 27th ACM SIGKDD Conference on Knowledge Discovery \& Data Mining}}. \bibinfo{pages}{3865--3874}.
\newblock


\bibitem[Ye et~al\mbox{.}(2021)]%
        {ye2021tr}
\bibfield{author}{\bibinfo{person}{Deming Ye}, \bibinfo{person}{Yankai Lin}, \bibinfo{person}{Yufei Huang}, {and} \bibinfo{person}{Maosong Sun}.} \bibinfo{year}{2021}\natexlab{}.
\newblock \showarticletitle{Tr-bert: Dynamic token reduction for accelerating bert inference}.
\newblock \bibinfo{journal}{\emph{arXiv preprint arXiv:2105.11618}} (\bibinfo{year}{2021}).
\newblock


\bibitem[Ye et~al\mbox{.}(2025)]%
        {Ye_2025_CVPR}
\bibfield{author}{\bibinfo{person}{Xubing Ye}, \bibinfo{person}{Yukang Gan}, \bibinfo{person}{Yixiao Ge}, \bibinfo{person}{Xiao-Ping Zhang}, {and} \bibinfo{person}{Yansong Tang}.} \bibinfo{year}{2025}\natexlab{}.
\newblock \showarticletitle{ATP-LLaVA: Adaptive Token Pruning for Large Vision Language Models}. In \bibinfo{booktitle}{\emph{Proceedings of the IEEE/CVF Conference on Computer Vision and Pattern Recognition (CVPR)}}. \bibinfo{pages}{24972--24982}.
\newblock


\bibitem[Yin et~al\mbox{.}(2024)]%
        {yin2024device}
\bibfield{author}{\bibinfo{person}{Hongzhi Yin}, \bibinfo{person}{Liang Qu}, \bibinfo{person}{Tong Chen}, \bibinfo{person}{Wei Yuan}, \bibinfo{person}{Ruiqi Zheng}, \bibinfo{person}{Jing Long}, \bibinfo{person}{Xin Xia}, \bibinfo{person}{Yuhui Shi}, {and} \bibinfo{person}{Chengqi Zhang}.} \bibinfo{year}{2024}\natexlab{}.
\newblock \showarticletitle{On-device recommender systems: A comprehensive survey}.
\newblock \bibinfo{journal}{\emph{arXiv preprint arXiv:2401.11441}} (\bibinfo{year}{2024}).
\newblock


\bibitem[Yu et~al\mbox{.}(2026)]%
        {thinkrec2026}
\bibfield{author}{\bibinfo{person}{Qihang Yu}, \bibinfo{person}{Kairui Fu}, \bibinfo{person}{Zheqi Lv}, \bibinfo{person}{Shengyu Zhang}, \bibinfo{person}{Xinhui Wu}, \bibinfo{person}{Chen Lin}, \bibinfo{person}{Feng Wei}, \bibinfo{person}{Bo Zheng}, {and} \bibinfo{person}{Fei Wu}.} \bibinfo{year}{2026}\natexlab{}.
\newblock \showarticletitle{{ThinkRec}: Thinking-based Recommendation via {LLM}}. In \bibinfo{booktitle}{\emph{Proceedings of the ACM Web Conference 2026}}. \bibinfo{pages}{5698--5709}.
\newblock


\bibitem[Zhan et~al\mbox{.}(2026)]%
        {slide2026}
\bibfield{author}{\bibinfo{person}{Tianyu Zhan}, \bibinfo{person}{Gui Ling}, \bibinfo{person}{Ran Tao}, \bibinfo{person}{Kunhai Lin}, \bibinfo{person}{Tong Xiong}, \bibinfo{person}{Wang Yang}, \bibinfo{person}{Kaixuan Zhang}, \bibinfo{person}{Zhihong Chen}, \bibinfo{person}{Yuliang Yan}, \bibinfo{person}{Dan Ou}, \bibinfo{person}{Shengyu Zhang}, {and} \bibinfo{person}{Haihong Tang}.} \bibinfo{year}{2026}\natexlab{}.
\newblock \showarticletitle{{SLIDE}: Search-to-List Item Distillation for Single-Rollout Generative Retrieval}.
\newblock \bibinfo{journal}{\emph{arXiv preprint arXiv:2610.34764}} (\bibinfo{year}{2026}).
\newblock


\bibitem[Zhan et~al\mbox{.}(2025)]%
        {cocorrrec2025}
\bibfield{author}{\bibinfo{person}{Tianyu Zhan}, \bibinfo{person}{Shengyu Zhang}, \bibinfo{person}{Zheqi Lv}, \bibinfo{person}{Jieming Zhu}, \bibinfo{person}{Jiwei Li}, \bibinfo{person}{Fan Wu}, {and} \bibinfo{person}{Fei Wu}.} \bibinfo{year}{2025}\natexlab{}.
\newblock \showarticletitle{Device-Cloud Collaborative Correction for On-Device Recommendation}. In \bibinfo{booktitle}{\emph{Proceedings of the Thirty-Fourth International Joint Conference on Artificial Intelligence}}. \bibinfo{pages}{3615--3623}.
\newblock


\bibitem[Zhang et~al\mbox{.}(2025)]%
        {zhang2025efficient}
\bibfield{author}{\bibinfo{person}{Qiang Zhang}, \bibinfo{person}{Hanchao Yu}, \bibinfo{person}{Ivan Ji}, \bibinfo{person}{Chen Yuan}, \bibinfo{person}{Yi Zhang}, \bibinfo{person}{Chihuang Liu}, \bibinfo{person}{Xiaolong Wang}, \bibinfo{person}{Christopher~E Lambert}, \bibinfo{person}{Ren Chen}, \bibinfo{person}{Chen Kovacs}, {et~al\mbox{.}}} \bibinfo{year}{2025}\natexlab{}.
\newblock \showarticletitle{Efficient Sequential Recommendation for Long Term User Interest Via Personalization}. In \bibinfo{booktitle}{\emph{2025 IEEE International Conference on Data Mining (ICDM)}}. IEEE, \bibinfo{pages}{913--922}.
\newblock


\bibitem[Zhao et~al\mbox{.}(2019)]%
        {zhao2019leveraging}
\bibfield{author}{\bibinfo{person}{Wei Zhao}, \bibinfo{person}{Benyou Wang}, \bibinfo{person}{Min Yang}, \bibinfo{person}{Jianbo Ye}, \bibinfo{person}{Zhou Zhao}, \bibinfo{person}{Xiaojun Chen}, {and} \bibinfo{person}{Ying Shen}.} \bibinfo{year}{2019}\natexlab{}.
\newblock \showarticletitle{Leveraging long and short-term information in content-aware movie recommendation via adversarial training}.
\newblock \bibinfo{journal}{\emph{IEEE transactions on cybernetics}} \bibinfo{volume}{50}, \bibinfo{number}{11} (\bibinfo{year}{2019}), \bibinfo{pages}{4680--4693}.
\newblock


\bibitem[Zhou et~al\mbox{.}(2018)]%
        {zhou2018deep}
\bibfield{author}{\bibinfo{person}{Guorui Zhou}, \bibinfo{person}{Xiaoqiang Zhu}, \bibinfo{person}{Chenru Song}, \bibinfo{person}{Ying Fan}, \bibinfo{person}{Han Zhu}, \bibinfo{person}{Xiao Ma}, \bibinfo{person}{Yanghui Yan}, \bibinfo{person}{Junqi Jin}, \bibinfo{person}{Han Li}, {and} \bibinfo{person}{Kun Gai}.} \bibinfo{year}{2018}\natexlab{}.
\newblock \showarticletitle{Deep interest network for click-through rate prediction}. In \bibinfo{booktitle}{\emph{Proceedings of the 24th ACM SIGKDD international conference on knowledge discovery \& data mining}}. \bibinfo{pages}{1059--1068}.
\newblock


\bibitem[Zhou et~al\mbox{.}(2020)]%
        {zhou2020s3}
\bibfield{author}{\bibinfo{person}{Kun Zhou}, \bibinfo{person}{Hui Wang}, \bibinfo{person}{Wayne~Xin Zhao}, \bibinfo{person}{Yutao Zhu}, \bibinfo{person}{Sirui Wang}, \bibinfo{person}{Fuzheng Zhang}, \bibinfo{person}{Zhongyuan Wang}, {and} \bibinfo{person}{Ji-Rong Wen}.} \bibinfo{year}{2020}\natexlab{}.
\newblock \showarticletitle{S3-rec: Self-supervised learning for sequential recommendation with mutual information maximization}. In \bibinfo{booktitle}{\emph{Proceedings of the 29th ACM international conference on information \& knowledge management}}. \bibinfo{pages}{1893--1902}.
\newblock


\bibitem[Zhu et~al\mbox{.}(2019)]%
        {DTCDR}
\bibfield{author}{\bibinfo{person}{Feng Zhu}, \bibinfo{person}{Chaochao Chen}, \bibinfo{person}{Yan Wang}, \bibinfo{person}{Guanfeng Liu}, {and} \bibinfo{person}{Xiaolin Zheng}.} \bibinfo{year}{2019}\natexlab{}.
\newblock \showarticletitle{DTCDR: A Framework for Dual-Target Cross-Domain Recommendation}. In \bibinfo{booktitle}{\emph{Proceedings of the 28th ACM International Conference on Information and Knowledge Management}} (Beijing, China) \emph{(\bibinfo{series}{CIKM '19})}. \bibinfo{publisher}{Association for Computing Machinery}, \bibinfo{address}{New York, NY, USA}, \bibinfo{pages}{1533–1542}.
\newblock
\showISBNx{9781450369763}
\href{https://doi.org/10.1145/3357384.3357992}{doi:\nolinkurl{10.1145/3357384.3357992}}


\end{thebibliography}

\clearpage
\setcounter{section}{0}
\renewcommand{\thesection}{S\arabic{section}}
\setcounter{figure}{0}
\renewcommand{\thefigure}{S\arabic{figure}}
\setcounter{algorithm}{0}
\renewcommand{\thealgorithm}{S\arabic{algorithm}}
\section*{Supplementary Material}

\section{Loss Coefficient Sensitivity Analysis}
\label{supp:loss_sensitivity}

\begin{figure}[h]
    \centering
    \includegraphics[width=1\linewidth]{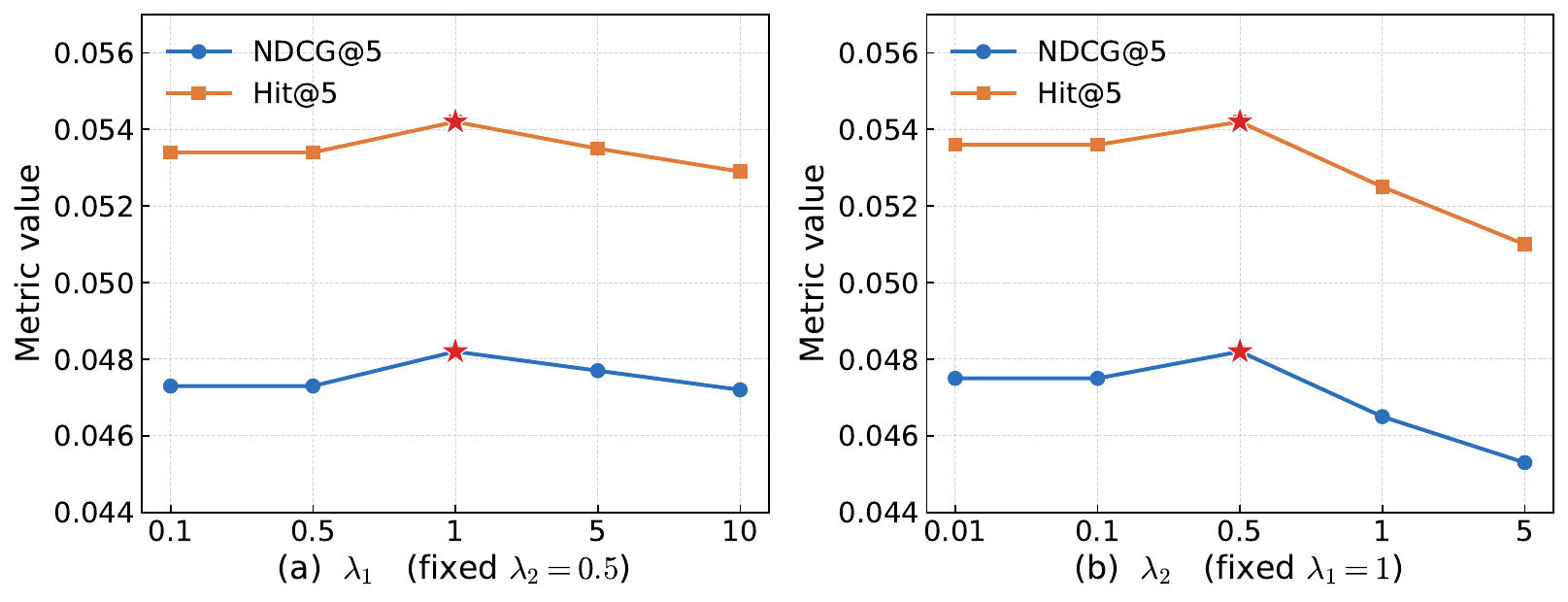}
    \vspace{-1mm}
    \caption{Sensitivity of NDCG@5 and Hit@5 to loss coefficients $\lambda_1$ and $\lambda_2$ on BSARec with Amazon-CD. Stars mark the default setting used in the main experiments.}
    \vspace{-1mm}
    \label{fig:hyper_analysis}
\end{figure}

We additionally examine the sensitivity of the two coefficients in the training objective (Eq.~\ref{eq:total_loss}) on BSARec with the Amazon-CD dataset. Figure~\ref{fig:hyper_analysis} reports NDCG@5 and Hit@5 while varying one coefficient and fixing the other at its default value.

Among the evaluated settings, the best observed configuration is $\lambda_1=1$ and $\lambda_2=0.5$. NDCG@5 varies by at most 2.1\% as $\lambda_1$ ranges from 0.1 to 10 and by at most 6.4\% as $\lambda_2$ ranges from 0.01 to 5. The smaller variation for $\lambda_1$ and the moderate variation for $\lambda_2$ show that the default is not a single sharply tuned point, while also indicating that performance is more sensitive to $\lambda_2$. Overall, the results support robustness across the evaluated coefficient ranges without implying complete invariance to their choice.

\section{Pseudocode}

\begin{algorithm}[b]
\begin{flushleft}
\caption{Pseudocode of RACE}

\textbf{Module 1:}~\colorbox{gray!30}{$\rhd$~\emph{Early-Layer Feature Encoding}}
\resizebox{0.46\textwidth}{!}{
    \begin{tcolorbox}[sharp corners, colframe=gray!80!white, colback=white, boxrule=0.5mm, left=0pt, right=3pt, top=1.5pt, bottom=0.5pt, boxsep=8pt]
    \begin{algorithmic}
    \State \textbf{Function}: Encode the historical sequence through the 0-th Transformer block $\mathcal{F}^{(0)}$ and repurpose the last valid token representation as a dynamic user embedding $\mathbf{c}$.
    \State \textbf{Input}: Historical Sequence $S^u$
    \State \textbf{Output}: Early-layer Features $\mathbf{H}^{(0)}$, Dynamic User Embedding $\mathbf{c}$
    \end{algorithmic}
    \end{tcolorbox}
    }
\textbf{Module 2:}~\colorbox{gray!30}{$\rhd$~\emph{Self-Supervised Threshold Prediction}}
    \resizebox{0.46\textwidth}{!}{
    \begin{tcolorbox}[sharp corners, colframe=gray!80!white, colback=white, boxrule=0.5mm, left=0pt, right=3pt, top=1.5pt, bottom=0.5pt, boxsep=8pt]
    \begin{algorithmic}
    \State \textbf{Function}: Anticipate instance-specific pruning ratios from the dynamic user embedding via a lightweight predictor $g_{\phi}$, supervised by the ideal threshold $\tau^*$ from Module 6.
    \State \textbf{Input}: Dynamic User Embedding $\mathbf{c}$
    \State \textbf{Output}: Channel Pruning Ratio $\tau$, Sequence Pruning Ratio $\tau'$
    \end{algorithmic}
    \end{tcolorbox}
    }
\textbf{Module 3:}~\colorbox{gray!30}{$\rhd$~\emph{Interest-Compatible Subnet Identification}}
    \resizebox{0.46\textwidth}{!}{
    \begin{tcolorbox}[sharp corners, colframe=gray!80!white, colback=white, boxrule=0.5mm, left=0pt, right=3pt, top=1.5pt, bottom=0.5pt, boxsep=8pt]
    \begin{algorithmic}
    \State \textbf{Function}: Analyze FFN channel sensitivities via the hypernetwork-style Sensitivity Extractor and carve out an instance-specific subnet according to channel pruning ratio $\tau$.
    \State \textbf{Input}: Hidden States $\mathbf{H}^{(l)}$, Channel Pruning Ratio $\tau$
    \State \textbf{Output}: Channel Mask $\mathbf{m}^{(l)}$
    \end{algorithmic}
    \end{tcolorbox}
    }
\textbf{Module 4:}~\colorbox{gray!30}{$\rhd$~\emph{Progressive Sequence Filtering}}
    \resizebox{0.46\textwidth}{!}{
    \begin{tcolorbox}[sharp corners, colframe=gray!80!white, colback=white, boxrule=0.5mm, left=0pt, right=3pt, top=1.5pt, bottom=0.5pt, boxsep=8pt]
    \begin{algorithmic}
    \State \textbf{Function}: Compute the cumulative importance score $s_t$ block by block and retain the top-$n_l^{\mathrm{keep}}$ items, where $n_l^{\mathrm{keep}}=\lceil |S^{(l)}|(1-\tau')\rceil$.
    \State \textbf{Input}: Attention Map $\bar{\mathbf{A}}^{(l)}$, Sequence Pruning Ratio $\tau'$
    \State \textbf{Output}: Retained Token Set $S^{(l+1)}$
    \end{algorithmic}
    \end{tcolorbox}
    }
\textbf{Module 5:}~\colorbox{gray!30}{$\rhd$~\emph{Capacity-Adaptive Inference}}
    \resizebox{0.46\textwidth}{!}{
    \begin{tcolorbox}[sharp corners, colframe=gray!80!white, colback=white, boxrule=0.5mm, left=0pt, right=3pt, top=1.5pt, bottom=0.5pt, boxsep=8pt]
    \begin{algorithmic}
    \State \textbf{Function}: Achieve capacity-adaptive inference for model and data, skipping pruned channels and filtered sequence block by block to generate the recommendation list.
    \State \textbf{Input}: Hidden States $\mathbf{H}^{(l)}$, Channel Mask $\mathbf{m}^{(l)}$, Retained Item Set $S^{(l+1)}$
    \State \textbf{Output}: Top-$k$ Recommendation List
    \end{algorithmic}
    \end{tcolorbox}
    }
\textbf{Module 6:}~\colorbox{gray!30}{$\rhd$~\emph{List-Wise Ranking Uncertainty}}~\textit{\small(training only)}
    \resizebox{0.46\textwidth}{!}{
    \begin{tcolorbox}[sharp corners, colframe=gray!80!white, colback=white, boxrule=0.5mm, left=0pt, right=3pt, top=1.5pt, bottom=0.1pt, boxsep=8pt]
    \begin{algorithmic}
    \State \textbf{Function}: Quantify ranking confidence via list-wise ranking uncertainty calculations over top-$K$ predictions and derive the pseudo-label $\tau^*$ to supervise Module 2.
    \State \textbf{Input}: Top-$K$ Predictive Mean $\boldsymbol{\mu}_{\mathcal{K}}$, Variance $\boldsymbol{\sigma}^2_{\mathcal{K}}$
    \State \textbf{Output}: Ranking Uncertainty $\mathcal{U}_{\text{rank}}$, Ideal Channel Pruning Ratio $\tau^*$
    \end{algorithmic}
    \end{tcolorbox}
    }
\label{Pseudocode}
\end{flushleft}
\end{algorithm}

We place the pseudocode of RACE in Algorithm~\ref{Pseudocode}. RACE can be summarized as six modules: Modules~1--5 form the inference pipeline that produces the Top-$k$ list under instance-specific channel and sequence pruning, while Module~6 runs only during training and supervises the threshold predictor (Module~2) via the pseudo-label $\tau^*$ derived from $\mathcal{U}_{\text{rank}}$. Together, the six modules realize an instance-specific, uncertainty-guided slimming pipeline that adaptively allocates model and input capacity to each instance. Here, $\tau$ and $\tau'$ denote channel and sequence pruning ratios, respectively, and larger values indicate more aggressive pruning.

RACE anticipates the list-wise uncertainty from early-layer features and decides the pruning ratios $\tau$ and $\tau'$ accordingly in Module~2, determining how aggressively each instance is slimmed in subsequent modules. For a hard instance with high anticipated ranking uncertainty, Module~2 outputs smaller ratios, leading Module~3 to retain more FFN channels in each Transformer block and Module~4 to keep more interaction items as the sequence propagates across blocks. Conversely, for an easy instance, larger thresholds drive more aggressive channel deactivation and sequence filtering, reducing inference cost more substantially while still delivering personalized and accurate recommendations.

\end{document}